\documentclass[superscriptaddress, amsmath, nofootinbib, twocolumn]{revtex4-1}

\usepackage{graphicx}
\usepackage{xcolor}
\usepackage{orcidlink}
\usepackage{enumerate}
\usepackage{enumitem}
\usepackage{ulem}
\usepackage{afterpage}
\newcounter{count}

\definecolor{linkcolor}{rgb}{0.0,0.3,0.5}
\definecolor{linkcolor}{rgb}{0.0,0.3,0.5}
\definecolor{mypurple}{RGB}{143, 116, 210}

\newcommand*{\defeq}{\mathrel{\vcenter{\baselineskip0.5ex \lineskiplimit0pt
                     \hbox{\scriptsize.}\hbox{\scriptsize.}}}
                     =}
\newcommand*{\invdefeq}{=\mathrel{\vcenter{\baselineskip0.5ex \lineskiplimit0pt
                     \hbox{\scriptsize.}\hbox{\scriptsize.}}}
                     }
\newcommand{\iu}{\mathrm{i}\mkern1mu}
\newcommand{\du}{\mathrm{d}}
\newcommand{\phicen}{|\phi_{\mathrm{c}}|}

\newcommand{\mpl}{M_{\rm Pl}}

\graphicspath{{Figures/}}

\newcommand{\cambridge}{Centre for Theoretical Cosmology, Department
of Applied Mathematics and Theoretical Physics, University of
Cambridge, Wilberforce Road, Cambridge CB3 0WA, United Kingdom}
\newcommand{\perimeter}{Perimeter Institute for Theoretical Physics, Waterloo, Ontario N2L 2Y5, Canada}
\newcommand{\sbu}{Institute for Advanced Computational Science,
Stony Brook University, Stony Brook, NY 11794 USA}
\newcommand{\kcl}{Theoretical Particle Physics and Cosmology Group,
Physics Department, Kings College London, Strand, London WC2R 2LS,
United Kingdom} 
\newcommand{\sapienza}{Dipartimento di Fisica, Sapienza Università
di Roma, Piazzale Aldo Moro 5, 00185, Roma, Italia} 
 \newcommand{\jhu}{Department of Physics and
Astronomy, Johns Hopkins University, 3400 North Charles Street,
Baltimore, Maryland 21218, USA} \newcommand{\caltech}{TAPIR 350-17,
Caltech, 1200 E. California Boulevard, Pasadena, California 91125,
USA}

\begin{document}
{\hfill KCL-TH-2024-59}

\title{Hair is complicated: Gravitational waves from stable and unstable boson-star mergers}

\author{Bo-Xuan Ge
\orcidlink{0000-0003-0738-3473}}
\email{bo-xuan.ge@kcl.ac.uk}
\affiliation{\kcl}
\author{Eugene A. Lim
\orcidlink{0000-0002-6227-9540}}
\email{eugene.a.lim@gmail.com}
\affiliation{\kcl}
\author{Ulrich Sperhake
\orcidlink{0000-0002-3134-7088}}
\affiliation{\cambridge}
\affiliation{\jhu}
\affiliation{\caltech}
\author{Tamara Evstafyeva
\orcidlink{0000-0002-2818-701X}}
\email{te307@cam.ac.uk}
\affiliation{\cambridge}
\affiliation{\perimeter}
\author{Daniela Cors
\orcidlink{0000-0002-0520-2600}}
\email{dc889@cam.ac.uk}
\affiliation{\cambridge}
\author{Eloy de Jong
\orcidlink{0000-0002-4505-0808}}
\email{eloydejong93@gmail.com}
\affiliation{\kcl}
\author{Robin Croft
\orcidlink{0000-0002-1236-6566}}
\email{robin.croft@uniroma1.it}
\affiliation{\sapienza}
\author{Thomas Helfer
\orcidlink{0000-0001-6880-1005}}
\email{thomashelfer@live.de}
\affiliation{\sbu}

\begin{abstract}
   We explore the gravitational-wave emission from head-on collisions of equal-mass solitonic boson-star binaries from simulations spanning a two-dimensional parameter space, consisting of the central scalar-field amplitude of the stars and the solitonic potential parameter.
   We report the gravitational-wave energies emitted by boson-star
   binaries which, due to their combination of moderately high
   compactness with significant deformability, we often find to be
   louder by up to an order of magnitude than analogous black-hole 
   collisions.
   The dependence of the radiated energy
   on the boson-star parameters exhibits striking needle-sharp
   features and discontinuous jumps to the value emitted by
   black-hole binaries. We explain these features in terms of
   the solitonic potential and the stability properties of the
   respective individual stars.
\end{abstract}
\maketitle

\section{Introduction}
\label{sec:intro}
Over the past 50 years, compact objects have acquired a center-stage
role in many areas of astrophysics and fundamental physics.  Neutron
stars (NSs) \cite{Hewish:1968bj,Shapiro1983} form the end product
of the evolution of moderately massive stars and, through binary
coalescence, give rise to short gamma-ray bursts
\cite{Berger:2013jza,LIGOScientific:2017ync}. Supermassive BHs
residing at the center of most (if not all) galaxies are intricately
related to their host dynamics and evolution \cite{Cattaneo:2009ub}
and can drive some of the most luminous objects in the universe
\cite{Rees:1984si}. NSs and stellar-mass BHs, when accreting matter
from a binary companion, manifest themselves as bright X-ray sources
\cite{Reig:2011zp} and so-called {\it primordial} BHs are hypothesized
as possible remnants of the early stages of cosmological evolution
\cite{Escriva:2022duf}. At a fundamental level, NSs represent unique
laboratories to study the equation of state of matter at super-nuclear
densities \cite{Ozel:2016oaf,LIGOScientific:2018cki,Lattimer:2021emm}
while BHs, through their rotation rates, may provide us with
information about matter fields beyond the standard model of particle
physics \cite{Cardoso:2014uka,Barack:2018yly,Cardoso:2018tly}. Both,
NSs and BHs enable us to test general relativity in the strong-field
regime \cite{Berti:2015itd,LIGOScientific:2021sio} and, as mathematical
solutions, have been essential in deepening our understanding of
Einstein's theory of general relativity; the Schwarzschild solution
in Kruskal-Szekeres coordinates or the Buchdahl limit are prominent
points in case.  On the observational side, the exploration of
compact objects has received a tremendous boost from the direct
detection of gravitational waves (GWs) by the Laser Interferometer
Gravitational-Wave Observatory (LIGO) and Virgo
\cite{Abbott:2016blz,KAGRA:2021vkt}.

One of the most exciting questions in this context is the very
nature of compact objects and whether these encompass {\it exotic}
members composed of matter beyond the standard model of particle
physics. Exotic compact objects may furthermore constitute a significant
fraction of the enigmatic {\it dark matter} content of the universe,
either themselves or as localized concentrations indicative of
fundamental fields permeating spacetime on grander scales. While a
plethora of experiments has been searching for dark-matter candidate
fields \cite{Feng:2010gw,Kahlhoefer:2017dnp,Cebrian:2022brv}, their
capacity to form gravitationally bound macroscopic concentrations
or ``stars'' offers the unique opportunity to study them as
compact-binary sources of GWs. Such an observational effort, however,
requires a detailed theoretical understanding and precision modeling
of exotic compact binaries for two main reasons. First, the canonical
approach to analyze observational data from GW detectors is based
on Bayesian statistics and matched filtering techniques employing
extensive theoretical GW template banks
\cite{Jaranowski:2005hz,Maggiore:2007ulw}. Second, as evidenced in
Ref.~\cite{Evstafyeva:2024qvp} for the case of boson stars and the
sensitivity of present GW detectors, the GW signals generated by
the inspiral and coalescence of exotic compact binaries can exhibit
significant degeneracy with those from black-hole (BH) binaries.
Breaking this degeneracy through the identification of characteristic
signatures from exotic compact objects will be of great benefit in
our GW guided quest for new physics.

The main goal of this paper is to explore in depth specific
characteristics of the GW emission from stars composed of a complex
scalar field. This class of compact objects, henceforth referred
to simply as {\it boson stars} (BSs) (see
e.g.~Refs.~\cite{Liebling:2012fv, Visinelli:2021uve, Bezares:2024btu} for reviews),
combines a deceptive simplicity regarding their mathematical
foundation with a remarkably rich phenomenology of their dynamics
and radiative properties. As such, BSs are not only interesting in
their own right, but also provide an excellent proxy to capture
main effects for a wider range of compact objects. Put another way,
should the analysis of a future GW event exhibit statistically
significant preference for BS templates relative to BH ones, the
impact would be substantial whether or not the source eventually
turns out to be a BS or another exotic source as for example cosmic
strings \cite{Aurrekoetxea:2023vtp} or compact stars composed of
higher-spin fields \cite{CalderonBustillo:2020fyi}.

The concept of BSs first appeared almost 60 years ago thanks to the
pioneering work by Kaup \cite{Kaup:1968zz}, whilst their dynamics
in binary systems have been studied numerically with increasing
intensity only for about 20 years. These studies have already
uncovered a considerable amount of valuable insight on the properties
of these exotic objects
by focusing on binaries consisting of rotating stars
\cite{Siemonsen:2020hcg, Sanchis-Gual:2019ljs, Yoshida:1997qf,
Mielke:2016war, Kleihaus:2005me, DiGiovanni:2020ror, Dmitriev:2021utv, Cardoso:2007az, Lai:2004fw}, BSs with light rings \cite{Siemonsen:2024snb,
Cunha:2017qtt, Cardoso:2016rao, Cunha:2022gde}, equal- \cite{Helfer:2021brt, Croft:2022bxq,
Palenzuela:2017kcg, Atteneder:2023pge, Bezares:2017mzk,
Sanchis-Gual:2020mzb, Palenzuela:2007dm, Mundim:2010hi} and
unequal-mass binaries \cite{Evstafyeva:2022bpr, Siemonsen:2023age,
Bezares:2022obu}, hybrid BS-NS~\cite{Dietrich:2018bvi} and
BS-BH~\cite{Clough:2018exo, Cardoso:2022vpj} binaries, multiple
scalar fields \cite{Bezares:2018qwa, Alcubierre:2018ahf,
Sanchis-Gual:2021edp, Jaramillo:2020rsv, Bernal:2009zy, Hawley:2002zn,
Guzman:2018evm}, real scalar fields (aka \textit{oscillatons})
\cite{Helfer:2018vtq, Brito:2015yfh, Widdicombe:2019woy,
Grandclement:2011wz} and vector fields (aka \textit{Proca stars})
\cite{Brito:2015pxa, Sanchis-Gual:2018oui, CalderonBustillo:2020fyi,
Herdeiro:2023wqf}. The BS parameter space offers ample avenues for
further exploration. For example, the impact of dephasing and the
vortex structure on the dynamics of inspiralling BS binaries has
only recently been studied in Ref.~\cite{Siemonsen:2023hko}.  Their
results demonstrate for the first time that a rotating BS can form
from the coalescence of two non-spinning BSs. Formation of these
objects are less well studied, but in \cite{Widdicombe:2018oeo},
it was shown that real-scalar stars can generically form from
the collapse of overdense regions. Similarly, thanks to increased
efforts to model exotic compact objects more extensively, a surrogate
model for equal-mass head-on collisions of Proca stars has been
constructed
in Ref.~\cite{Luna:2024kof}.

The merger dynamics of BS binaries have been found to exhibit a
(perhaps surprisingly) rich phenomenology driven by the specific
properties of the individual BSs.  For instance,
Ref.~\cite{Palenzuela:2006wp} observes that the GW emission of the
simulated head-on collisions of equal-mass mini BSs is quadrupole
dominated but with varying characteristics. On the one hand, in-phase
BSs merge into a single more massive star whereas, on the other
hand, binaries with maximally dephased scalar fields come close to
each other but remain separate, almost like two touching snooker balls. For
large initial momenta, the two mini BSs can even pass through each
other \cite{Choi:2009,Widdicombe:2019woy}, exhibiting a true solitonic nature. An even
more exotic fate of an inspiralling BS binary has been found in
Ref.~\cite{Siemonsen:2023age}, where an aligned-spin binary, instead
of merging, forms a binary of nonspinning BSs flung out away from
the center of mass.  Ultrarelativistic BS collisions were modeled
in Ref.~\cite{Choptuik:2009ww} and found to result in BH formation,
consistent with predictions by Thorne's hoop conjecture
\cite{Thorne:1972ji}.

Numerical explorations of BSs have thus demonstrated many interesting
features of their mergers and dynamics; however, our insight is
largely limited to isolated points or patches in the BS parameter
space leaving systematic explorations of the complete parameters
as a key challenge for complementary work.  The most comprehensive
study in this direction has, to our awareness, been performed for
Proca stars in Ref.~\cite{Sanchis-Gual:2022mkk}, covering a range
of frequency (i.e.~mass) ratios and phase offsets.  The central
goal of our work is to extend the parameter coverage to the potential
function for head-on collisions of equal-mass non-spinning scalar
BSs. For this purpose we simulate $\sim 10^4$ BS merger systems composed
of a scalar field of mass $m$ self-interacting through a solitonic
potential
\begin{equation}
  V(\phi) = m^2 |\phi|^2 \left(
  1 - 2\frac{|\phi|^2}{\sigma^2}
  \right)^2\,,
  \label{eq:Vsol}
\end{equation}
where $\sigma$ is the self-interaction parameter.
Specifically, we consider a two-dimensional hypersurface of the
parameter space spanned by $\sigma$ and the BS compactness controlled
through the central scalar-field amplitude $|\phi_{\rm c}|$. Across
this parameter range, we monitor the BS dynamics through infall and
merger, the nature of the remnant and, most importantly, the resulting
GW emission.

Throughout this work, we employ natural units where
$\hbar=1=c$ but keep the gravitational constant $G$ related
to the Planck mass by $M_{\rm Pl}=\sqrt{G}^{-1}$.

\section{Executive summary}
\label{sec:precis}
\begin{figure*}
  \centering
  \includegraphics[width=0.48\textwidth]{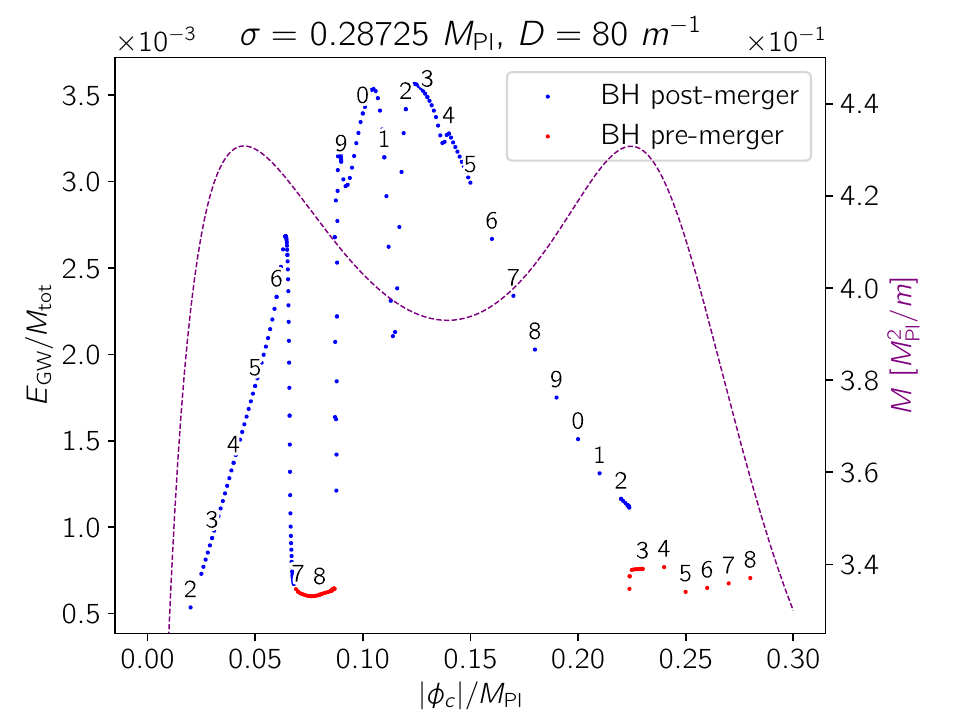}
  \label{oldfig:sigma=0.28725}
   \includegraphics[width=0.48\textwidth]{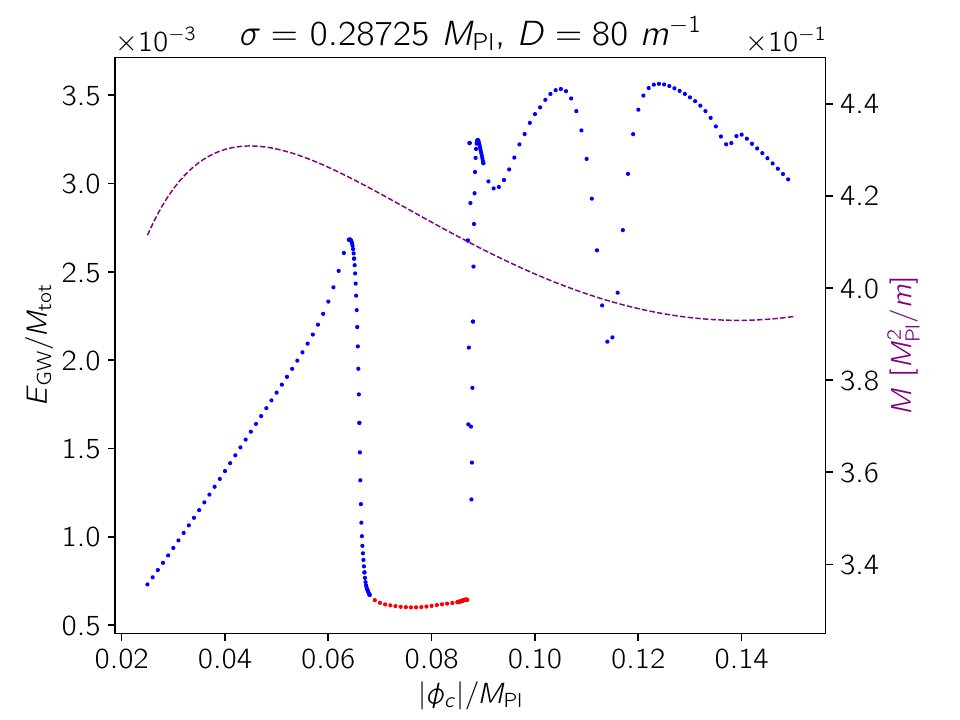}
  \label{fig:sigma=0.28725zoom}
  \caption{
  \textit{Left}: The GW energy of the head-on collision of two BSs
  with a solitonic potential (\ref{eq:Vsol}) for $\sigma=0.27825 M_{\rm{Pl}}$
  is shown as a function of the BSs' central scalar-field amplitude
  $|\phi_{\rm c}|$.
  Blue dots represent binaries completing their infall as
  BSs and forming a BH or BS
  at merger whereas red symbols denote systems where
  each BS collapses to a BH prior to merger.
  The numbers adjacent to the dots give the
  final digit of their $|\phi_{\rm c}|$ value.
  The purple line
  represents the mass curve $M(|\phi_{\rm c}|)$ for single spherical BSs
  with central scalar-field amplitude $|\phi_{\rm c}|$.
  \textit{Right}: Zoom into the region around the sharp variations
  of $E_{\rm GW}(|\phi_{\rm c}|)$.
  }
  \label{fig:appetizer}
\end{figure*}

The pr{\'e}cis
of our findings is presented in Fig.~\ref{fig:appetizer}
where we plot for $\sigma=0.28725\,M_{\rm Pl}$
the GW energy, normalized by the total BS mass
$M_{\rm tot}=M_1+M_2\invdefeq 2M$,
generated in the head-on collision of two BSs
with central scalar-field amplitude $|\phi_{\rm c}|$,
initial separation $80\,m^{-1}$ and initial velocity $v=0.1$
towards each other.
This figure demonstrates several remarkable features which we will
discuss and explain in physical terms in this paper.
\begin{list}{\rm{(\arabic{count})}}{\usecounter{count}
             \labelwidth0.5cm \leftmargin0.7cm \labelsep0.2cm \rightmargin0cm
             \parsep0.5ex plus0.2ex minus0.1ex \itemsep0ex plus0.2ex}
\item
\emph{BS collisions can be radiatively way more efficient than those
of BHs.} The GW energy released in the BS collisions reaches values
up to $E_{\rm GW}/M_{\rm tot}=0.0035$, about six times larger than
the corresponding number $6\times 10^{-4}$ observed in head-on
collisions of non-spinning equal-mass BHs; see e.g.~Fig.~4 in
Ref.~\cite{Sperhake:2019oaw} and also compare with the red
dots in our Fig.~\ref{fig:appetizer} where the BSs have collapsed
to BHs prior to merger. We will encounter yet larger energies when
we discuss other $\sigma$ values later on.  This effect was first
noted in Ref.~\cite{Helfer:2018vtq} in the context of self-gravitating
real-scalar-field oscillatons, suggesting that this is a generic
effect for mergers of compact objects composed of fundamental fields.
\item
\emph{There are discontinuities in the GW energy viewed as a function
of $|\phi_{\rm c}|$, i.e.~the BS compactness.} At $|\phi_{\rm
c}|/M_{\rm Pl}\gtrsim 0.08$, the energy $E_{\rm GW}$ abruptly jumps from the
BH value $E_{\rm GW}^{\rm BH}\approx 6\times 10^{-4}\,M_{\rm tot}$
(red dots) to about twice this value (the next right
blue dot). Between $|\phi_{\rm c}|/M_{\rm Pl}=0.22$ and
$0.23$, the function $E_{\rm GW}(|\phi_{\rm c}|)$ exhibits a
further discontinuity; even though this jump is relatively small,
it is well resolved by our numerics.
\item
Even away from these discontinuities, \emph{small differences in
the initial configuration lead to large, albeit continuous,
${\cal O}(1)$ differences in
the GW emission.} The functional dependence of the GW energy on
$|\phi_{\rm c}|$ exhibits sharp local extrema, most notably the local
minimum at $|\phi_{\rm c}|/M_{\rm Pl} \approx 0.115$; we will find
these sharp extrema -- maxima or minima -- to be even more prominent
for other $\sigma$ values and, henceforth, refer to them as {\it
needles}.
This observation suggests that it can be misleading to assume
similar BS collisions produce approximately equal output in GWs.
\item
We discern significant correlation -- overall positive for small
and negative for large $|\phi_{\rm c}|$ -- of the radiated energy
$E_{\rm GW}(|\phi_{\rm c}|)$ when compared to the mass function $M(|\phi_{\rm
c}|)$. We will show how this correlation can be explained in terms
of the shape of the potential $V(|\phi_{\rm c}|)$ and that
$E_{\rm GW}(|\phi_{\rm c}|)$ must be extremal near the
minimum of the mass function $M(|\phi_{\rm c}|)$
provided there exist stable BSs in a neighbourhood to the right
of this $|\phi_{\rm c}|$ value.
\end{list}

Our discussion of these effects is organized as follows.
In Sec.~\ref{sec:numerics} we summarize our computational setup and calibrate
the numerical uncertainty. The structure and stability of single BSs
and their solution branches is analyzed in Sec.~\ref{sec:single}
and in Sec.~\ref{sec:collisions}, we present the key findings of
our investigation of BS collisions and their GW emission. We discuss
our findings in Sec.~\ref{sec:discussion} and present some details
of the GW extraction in the appendix.

\section{Theory and computational framework}
\label{sec:numerics}
BSs composed of a single complex scalar field $\phi$ minimally
coupled to the spacetime metric $g_{\alpha\beta}$ of general
relativity are described by the Lagrangian
\begin{equation}
  \mathcal{L} = \frac{1}{16\pi G}R-\frac{1}{2} g^{\mu \nu}
  \nabla_\mu \bar{\phi} \nabla_\nu \phi - \frac{1}{2} V(\phi)\,,
  \label{eq:Lbs}
\end{equation}
where an overbar denotes the complex conjugate and $V$ is the
potential function given in our case by Eq.~(\ref{eq:Vsol}). Variation
of the action $S=\int \sqrt{-g}\mathcal{L}\du^4 x$ with respect to
the metric and scalar field results in the covariant equations
\begin{eqnarray}
  && G_{\alpha\beta} = 8\pi G \left\{
  \nabla_{(\alpha}\bar{\phi}\,\nabla_{\beta)}\phi
  -\frac{1}{2} \left[
  \nabla^{\mu}\bar{\phi}\,\nabla_{\mu}\phi+V(\phi)
  \right]
  \right\}\,,
  \nonumber \\
  && \nabla^{\mu}\nabla_{\mu} \phi = \frac{\du V}{\du |\phi|^2}\phi\,.
\end{eqnarray}
The calculation of spherically symmetric single-BS spacetimes is
most conveniently performed by decomposing the scalar field into
amplitude and phase according to
\begin{equation}
  \phi(t, r) = |\phi(r)| e^{\mathrm{i} \omega t},
  \label{eq:phidecomp}
\end{equation}
where $|\phi(r)|$ is the scalar-field amplitude and $\omega$ the
(constant positive) frequency. This ansatz can be generalized by
writing the phase factor as $e^{\iu(\epsilon \omega t + \delta
\vartheta)}$, where $\epsilon = \pm 1$ distinguishes BSs and anti-BSs,
and $\delta \vartheta$ denotes a phase shift. In this work, we
exclusively consider BS binaries without phase offset and, hence,
set $\epsilon=1$, $\delta \vartheta=0$. With the decomposition
(\ref{eq:phidecomp}), the Einstein-Klein-Gordon equations for a spherically
symmetric metric
\begin{equation}
  \du s^2 = -\alpha(r)^2\du t^2 + X(r)^2\du r^2 + r^2(\du \theta^2
  +\sin^2\theta \du \varphi^2)\,,
  \nonumber
\end{equation}
result in a coupled set of four first-order ordinary differential
equations which we solve with a shooting method as detailed in
Sec.~2.3 of Ref.~\cite{Helfer:2021brt}; cf.~also Appendix A of
Ref.~\cite{Evstafyeva:2023kfg} for an in-depth discussion of the
solution's asymptotic behaviour at infinity.

For BS binary evolutions, we construct initial data by combining
two such single-BS spacetimes using the {\it constant volume element}
technique of Refs.~\cite{Helfer:2018vtq,Helfer:2021brt,Evstafyeva:2022bpr};
cf.~also Ref.~\cite{Atteneder:2023pge}.
This technique results in a significant improvement in the
Hamiltonian constraint equation over the simple linear superposition
of the two BS solutions.

The resulting binaries are
characterized by the parameter $\sigma$ in the solitonic potential
(\ref{eq:Vsol}), the stars' central scalar-field amplitudes $|\phi_{\rm c}|$,
their initial separation $D$ and their velocities $v$. Our study
focuses on non-spinning equal-mass BS binaries, so that both stars
have equal $|\phi_{\rm c}|$ and, in the rest frame used for all our runs,
their initial velocity has equal magnitude with direction pointing
toward each other.  Unless specified otherwise, we fix $v=0.1$ and
$D=80\,m^{-1}$ which leaves us with two control parameters, $\sigma$
and $|\phi_{\rm c}|$.  In this paper, we present results for $\sigma/M_{\rm
Pl}=\{0.25, 0.28725, 0.3, 0.5\}$ as well as the mini-BS limit
$\sigma\rightarrow\infty$; these cases are fully representative of
the dynamics under consideration, but we note that further values
of $\sigma$ are discussed in Ref.\cite{Ge2024}.

\begin{figure}
  \includegraphics[width=0.45\textwidth]{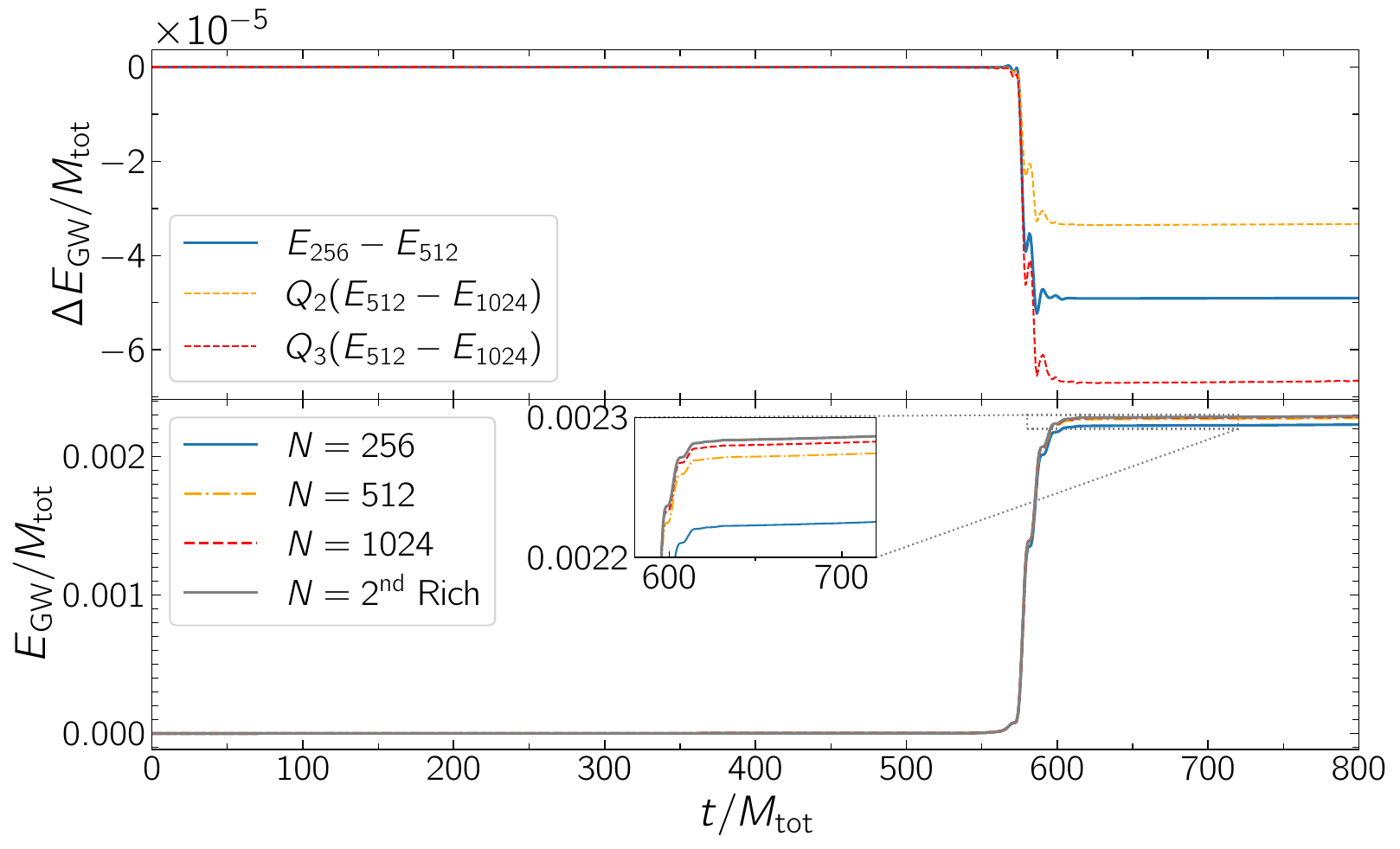}
  \caption{Convergence analysis of the GW energy for the binary
  BS configuration $\sigma=0.25\,M_{\rm Pl}$, $\phicen=0.18\,M_{\rm Pl}$.
  The top panel displays the differences in the energy as a function
  of time for resolutions
  $N=256$, $N=512$ and $N=1024$ points with the higher-resolution
  differences magnified by factors $Q_2=4$ and $Q_3=8$ expected
  for second and third-order convergence. The bottom panel
  shows the radiated energy as a function of time for all resolutions
  together with the result for second-order Richardson extrapolation.
  }
\label{fig:conv_s025_a018}
\end{figure}
We simulate the head-on collisions resulting from these initial
data using the {\sc grchombo} code \cite{Clough:2015sqa, Radia:2021smk,
Andrade:2021rbd} which evolves the Einstein equations in the CCZ4
formulation \cite{Alic:2011gg}.  Our specific version of the CCZ4
equations is given in Sec.~2 of Ref.~\cite{Radia:2021smk} and is
discretized using the {\it method of lines} with fourth-order finite
differencing in space and a fourth-order Runge-Kutta scheme in time.
{\sc grchombo} implements Berger-Rigoutsos adaptive-mesh refinement
provided by the {\sc chombo} library \cite{chombo} and facilitates
various tagging criteria for the grid generation as summarized in
Sec.~3 of Ref.~\cite{Radia:2021smk}.  In this paper, we utilize the
scalar field $|\phi|$ and the conformal factor $\chi$ for regridding
with their respective thresholds set to 0.3.
We exploit the inherent axisymmetry of our BS binary configurations
by reducing our computational domain from three to two spatial
dimensions using the {\it modified cartoon method} in the form
detailed in Ref.~\cite{Cook:2016soy}; cf.~also
\cite{Alcubierre:1999ab,Pretorius:2004jg} as well as Appendix
\ref{app:waveextraction} for our cartoon implementation of the wave
extraction.  This combination of mesh refinement and dimensional
reduction enables us to simulate approximately 10\,000 simulations
-- each utilizing two nodes (56 logical cores) and lasting approximately
four hours.
We employ a domain of length $512\,m^{-1}$
and a grid setup consisting of a nested set of $L=7$ refinement
levels with grid spacing $\du x=1/(32\,m)$
on the innermost level, increasing by a factor 2 consecutively on
each level further out.

For the calibration of our numerical uncertainty,
we have simulated a representative
BS binary configuration with $\sigma=0.25\,M_{\rm Pl}$ and
$\phicen=0.18\,M_{\rm Pl}$ using three resolutions, our default
$N=256$ as well as two higher resolutions $N=512$ and $N=1024$.
The result, displayed in Fig.~\ref{fig:conv_s025_a018},
demonstrates convergence between second and third order. We
conservatively estimate the discretization error by assuming
second-order convergence for the Richardson extrapolation and obtain
a relative error of about $2.5\,\%$ for $N=256$. By fitting the energy values obtained
at different extraction radii with a polynomial in $1/R_{\rm ex}$
truncated at second order, we find an additional uncertainty of 
$\sim 2\,\%$ at our fiducial $R_{\rm ex}=140\,m^{-1}$
resulting in a total error budget of $\sim 4.5\,\%$.
We have also monitored the constraint violations of our
simulations as described in Appendix \ref{app:ham}.

\begin{figure}
  \includegraphics[width=0.45\textwidth]{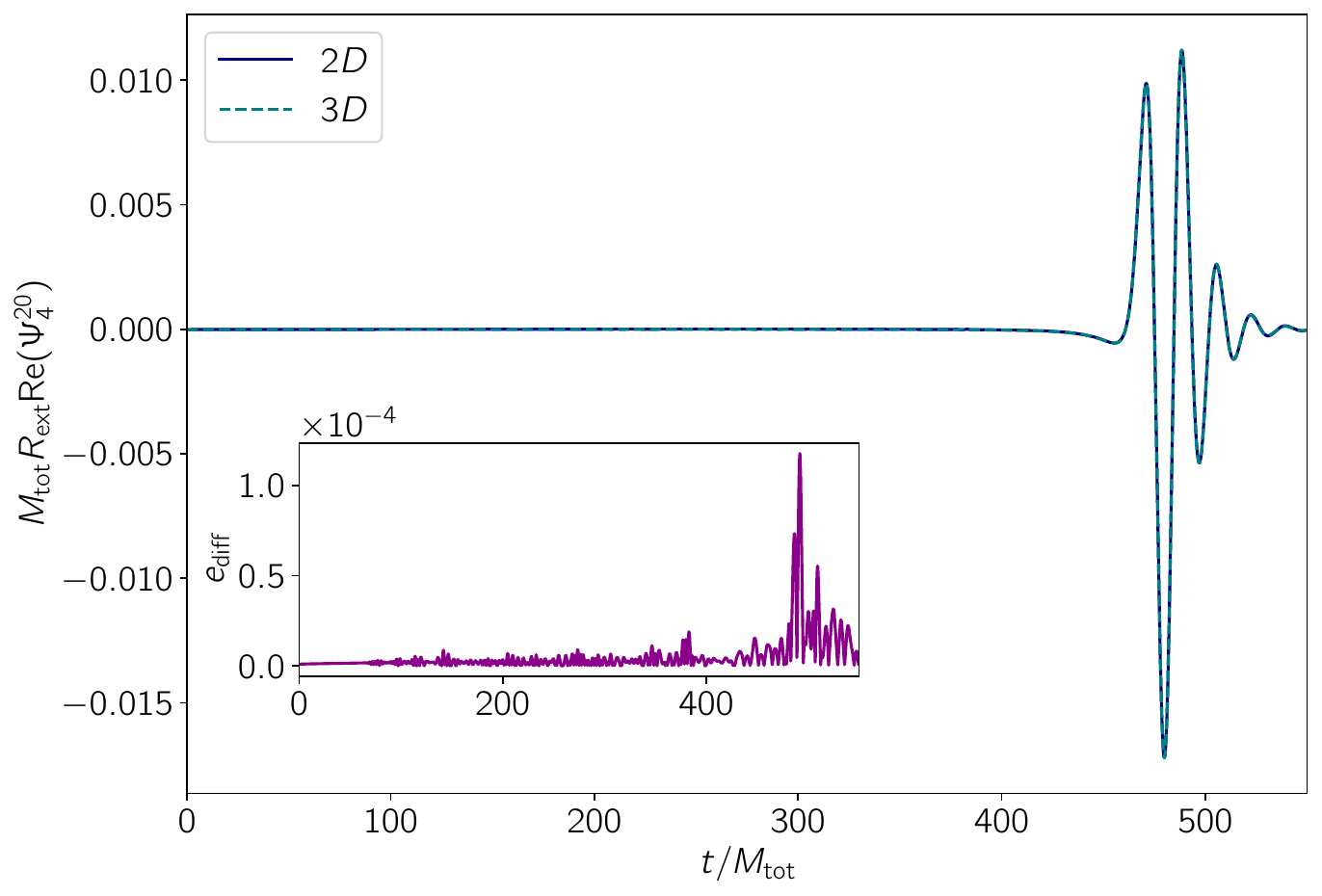}
  \caption{
  Comparison of the $(2,0)$ GW mode generated by the head-on collision
  of equal-mass BSs with $|\phi_{\rm{c}}| = 0.086 M_{\rm{Pl}}$, $\sigma = 0.35 M_{\rm{Pl}}$ using
  the 2D and 3D versions of the {\sc GRChombo} code. The inset shows the
  absolute difference $e_{\rm{diff}}$ between the two signals and
  demonstrates relative agreement $\lesssim 1\,\%$, well within the uncertainty of the simulations.
  }
\label{fig:3Dvs2DGW}
\end{figure}

As a final test, we compare in Fig.~\ref{fig:3Dvs2DGW} the GW signal
for a BS head-on collision for $\sigma=0.35\,M_{\rm Pl}$ and $|\phi_{\rm
c}|=0.086\,M_{\rm Pl}$ calculated with our modified
Cartoon approach with the result from a fully three-dimensional
computation. The results agree within about a percent, commensurate
with the uncertainty of the three-dimensional simulation
\cite{Helfer:2021brt}.

\section{Single boson stars}
\label{sec:single}

\subsection{Stationary boson stars}
\label{sec:single_stationary}
Many of the prominent features we observe in the dynamics and GW
emission of BS head-on collisions find their origin in the properties
of single BS spacetimes. We therefore start the discussion of our
results by recalling the essential characteristics of spherically
symmetric, stationary BS models. For a fixed scalar potential,
Eq.~(\ref{eq:Vsol}) in our case, the space of spherically symmetric
solutions is controlled by the central scalar-field amplitude
$|\phi_{\rm c}|$. More specifically, for each value $|\phi_{\rm
c}|$, there exists a countably infinite number of BS models
characterized by the number $n$ of zero crossings of the radial
profile $|\phi(r)|$: the {\it ground-state} BS with $n=0$ and the
{\it excited}\footnote{Not to be confused with ``perturbed'' or
``dynamic'' BSs which are stars evolving in time. Excited BSs, in
contrast, are equilibrium solutions merely having one or more zero
crossings in their radial scalar-field profile.
}
stars for $n=1,\,2,\,\ldots$\,. For a broad range of scalar potentials,
the latter have been found to be
intrinsically unstable to generic perturbations \cite{Balakrishna:1997ej},
but more recent studies have found
that sufficiently large self-interaction terms may cure such instabilities
\cite{Sanchis-Gual:2021phr,Brito:2023fwr}. In any case, excited
BSs will play no further role in this paper.

The remaining ground-state models represent a one-parameter family
of BSs with each member determined by its central scalar amplitude
$|\phi_{\rm c}|$. The solution space is conveniently displayed as
a curve $M(|\phi_{\rm c}|)$ in the {\it mass-amplitude}\footnote{
Alternatively, the solutions can be represented in a mass-radius
or $M(R)$ diagram.} plane where, analogous to the
Tolman-Oppenheimer-Volkoff \cite{Oppenheimer:1939ne,Tolman:1939jz}
solutions for spherically symmetric NSs, it divides into stable and
unstable branches \cite{Cook1994, Friedman1988, Harrison1965,
Straumann1984}.  These branches are delineated by local extrema of
the function $M(|\phi_{\rm c}|)$ where the time dependence of radial
peturbations changes from oscillatory to exponential
\cite{Lee:1988av,Gleiser:1988rq,Gleiser:1988ih, Jetzer:1988vr,
Jetzer:1989qp,
Jetzer:1989vs,Seidel:1990jh,Balakrishna:1997ej,Alcubierre:2003sx}. For
so-called {\it mini BSs}, where $\sigma \rightarrow \infty$, composed
of a non-interacting scalar field, for example, the $M(|\phi_{\rm
c}|)$ curve consists of one stable ($|\phi|<|\phi|_{\rm Kp}$) and
one unstable branch ($|\phi|>|\phi|_{\rm Kp}$) separated by the
maximum-mass or {\it Kaup limit} \cite{Kaup:1968zz} model with
$M=M_{\rm Kp}\approx 0.633\,M_{\rm Pl}^2/m$ for $|\phi_{\rm c}|
=|\phi|_{\rm Kp}\approx 0.0765\,M_{\rm Pl}$. For the solitonic
potential (\ref{eq:Vsol}), however, the branch structure becomes
more complicated; more precisely, it mirrors the more complicated
structure of the potential function $V(|\phi|)$ itself.

We can qualitatively understand this by considering the simplified
scenario of an approximately stationary and spatially uniform scalar
field. The energy of such a field is dominated by the potential
term $V(|\phi|)$ and we expect it to roll towards the stable
equilibrium at a local minimum of this function. Considering now the
solitonic potential (\ref{eq:Vsol}), we find three extrema, a local
maximum at $|\phi|=\sigma/\sqrt{6}$ and the two minima $|\phi|=0$
and $|\phi|=\sigma/\sqrt{2}$. The latter two correspond to stable
configurations while $|\phi|=\sigma/\sqrt{6}$ is unstable.  BSs
are, of course, {\it not} spatially uniform; instead their scalar
amplitude spans the entire range from $|\phi|=0$ at infinite radius
to $|\phi_{\rm c}|$ at the origin. And yet, let us take the analogy
as qualitative guidance for the moment, making sure to have a pinch
of salt at the ready.

Recalling from the above discussion of NSs and mini BSs that local
extrema in the $M(|\phi_{\rm c}|)$ curve typically\footnote{ The
rule is not rigorous; for example increasing $|\phi_{\rm c}|$ well
beyond the Kaup limit for mini BSs ultimately leads to further
extrema in $M(|\phi_{\rm c}|)$ for ultra-compact BSs, none of them
marking a return to stable BSs.  The presence or absence of a change
in stability depends on whether the static perturbation belongs to
the fundamental radial oscillation mode or an overtone
\cite{Gleiser:1988ih}.
}
mark the transition between stable and unstable models, stable BS
models will be located on rising slopes of $M(|\phi_{\rm c}|)$,
i.e.~have $\du M/\du |\phi_{\rm c}|>0$ while unstable models mainly
coincide with $\du M/\du |\phi_{\rm c}| <0$. Our crude analogy to
homogeneous scalar fields would then suggest that minima (maxima)
in $V(|\phi_{\rm c}|)$ coincide with rising (descending) slopes of
$M(|\phi_{\rm c}|)$.

\begin{figure}
  \centering
  \includegraphics[width=0.48\textwidth]{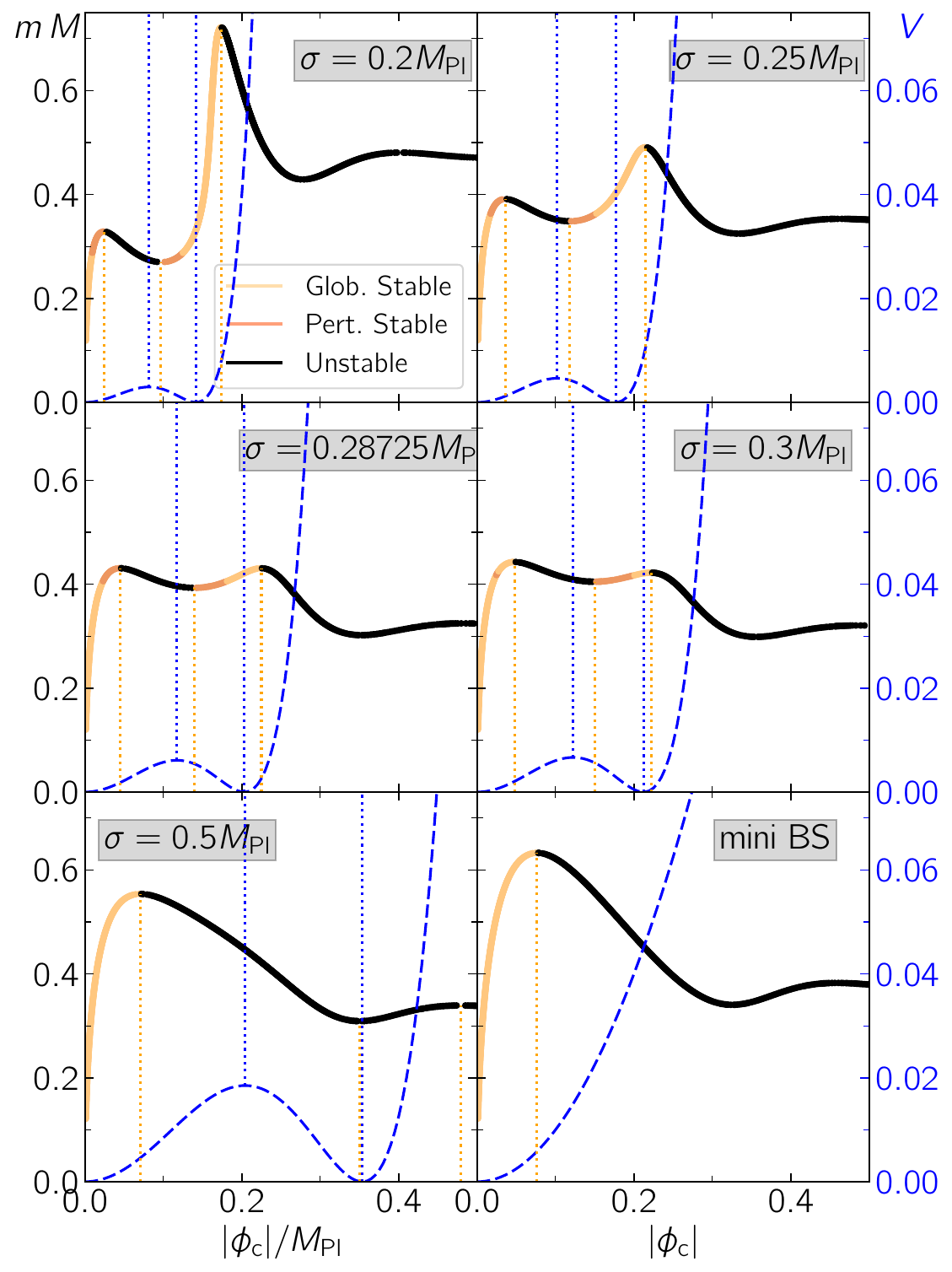} 
  \caption{
  $M(|\phi_{\rm c}|)$ diagrams for the one-parameter families of BSs obtained
  for $\sigma/M_{\rm Pl}=0.2$, $0.25$, $0.28725$, $0.3$, $0.5$ as
  well as the mini-BS limit $\sigma \rightarrow \infty$. For each
  one-parameter family of BS models, globally (perturbatively)
  stable models are displayed in light (dark) copper color while
  unstable stars are marked in black. The blue dashed curves,
  quantified in units of $m^2M_{\rm Pl}^2$
  on the right vertical axis, represent the respective
  potential functions (\ref{eq:Vsol}). The vertical dotted lines mark the
  extrema of the potential (blue) and the mass-amplitude curves (orange).
  }
  \label{fig:multistability}
\end{figure}

We illustrate this behaviour in Fig.~\ref{fig:multistability} for
five values $0.2 \le \sigma/M_{\rm Pl} \le 0.5$ and the mini-BS
limit $\sigma \rightarrow \infty$.  There we distinguish between
unstable BSs, perturbatively stable stars and globally stable BS
models, the latter defined by minimizing the binding energy
$M-\mathcal{N}$ over all BS models of equal Noether charge
$\mathcal{N}$.  We see in this figure that our association of maxima
in $V(|\phi_{\rm c}|)$ with unstable stars is borne out for all
potentials. Our expectation that minima in $V(|\phi_{\rm c}|)$
correspond to stable stars, on the other hand, is true for $\sigma
\lesssim 0.4\,M_{\rm Pl}$ but only holds for the true vacuum
$|\phi_{\rm c}|=0$ when $\sigma \gtrsim 0.4\,M_{\rm Pl}$. In this
latter regime, the false vacuum state $|\phi|=\sigma/\sqrt{2}$ has
shifted to rather large values corresponding to very compact stars
where gravitational effects dominate over scalar self interaction
and our simplified model expectedly starts showing cracks.

%
\begin{figure}
  \centering
  \includegraphics[width=0.5\textwidth]{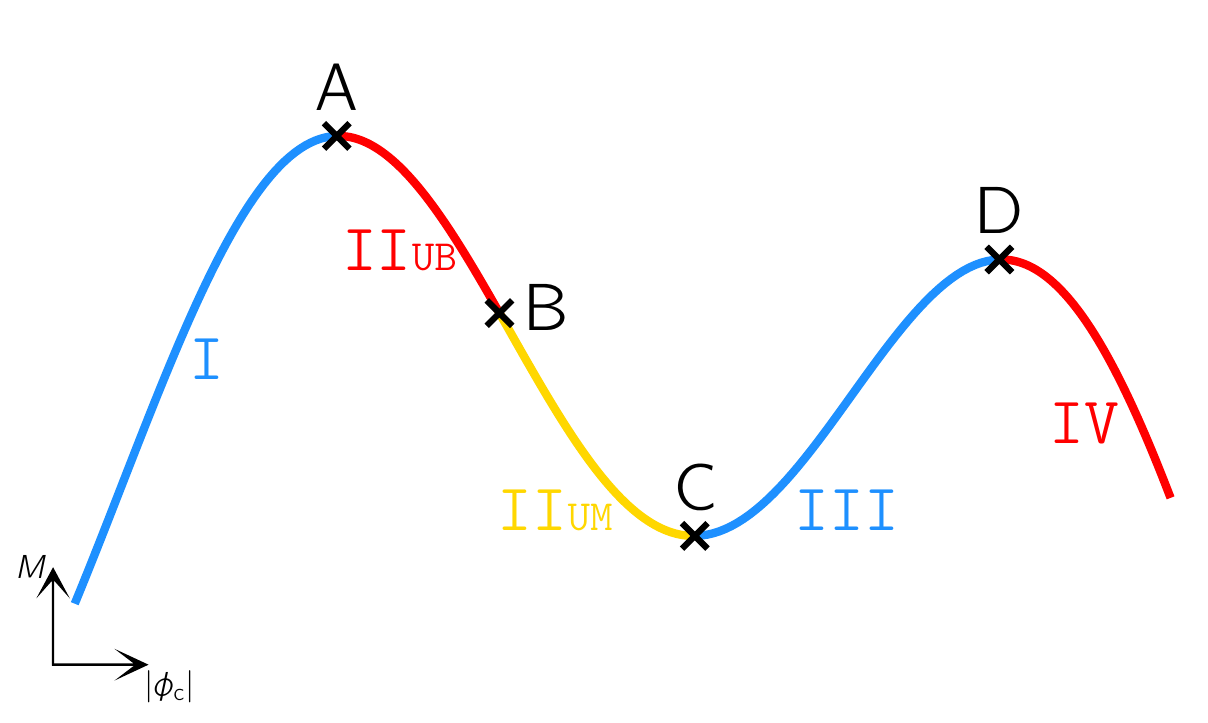}
  \caption{
  Schematic illustration of the $M(|\phi_{\rm c}|)$ curve and the
  resulting branches for stationary BSs. The local extrema naturally
  divide the $|\phi_{\rm c}|$ range into regions I and III with
  $\du M/\du |\phi_{\rm c}|>0$ and regions II and IV with $\du M/\du
  |\phi_{\rm c}|<0$. BSs of the former regions -- provided they are
  not extremely compact -- are stable (marked in blue) while the
  latter regions are exclusively comprised of unstable stars which,
  when perturbed, either collapse to BHs (red) or migrate to more
  compact stable BS configurations (yellow). We correspondingly
  call these branches \textbf{S} (stable), \textbf{UB} (unstable
  BH) and \textbf{UM} (unstable migrating).
  }
  \label{fig:stability_schematic}
\end{figure}

We summarize our observations by displaying schematically in
Fig.~\ref{fig:stability_schematic} the first three (in the sense
of $|\phi_{\rm c}|$ increasing from 0 towards positive values)
extrema of the $M(|\phi_{\rm c}|)$ curves computed in
Fig.~\ref{fig:multistability}.  These extrema naturally divide the
$|\phi_{\rm c}|$ parameter range into four regions, labeled I to
IV, where, alternatingly, $\du M/\du |\phi_{\rm c}|>0$ for odd and
$\du M/\du |\phi_{\rm c}|<0$ for even region number.  BS models in
region I are always stable whereas BS models in regions II and IV
are always unstable.  Models in region III are stable {\it provided}
they are not too compact; otherwise they are unstable. The structure
of the branches and their relative height furthermore suggest the
following possible fates for unstable stars perturbed out of
equilibrium.
\begin{list}{\rm{(\roman{count})}}{\usecounter{count}
             \labelwidth0.5cm \leftmargin0.7cm \labelsep0.2cm \rightmargin0cm
             \parsep0.5ex plus0.2ex minus0.1ex \itemsep0ex plus0.2ex}
\item
The BS evaporates with the entire scalar-field matter dispersing
to future timelike infinity $i^+$.
\item
The BS sheds part of its matter and migrates to a stable model with
lower mass.
\item
The BS migrates to a more compact stable configuration with equal
Noether charge.
\item
The BS collapses to a BH.
\end{list}
Roughly speaking we may consider this sequence of scenarios as
pointing in the direction from the scalar-dominated to the
gravity-dominated regime.  Empirically, as discussed in detail in
the next subsection, we exclusively observe cases (iii) and (iv);
we suspect that the shedding of matter or evaporation, while
theoretically possible, requires delicately fine tuned initial
perturbations corresponding to a bulk outward motion of the matter.
The possibility of migration to a more compact stable BS naturally
depends on the existence of such models and, therefore, on the
relative height of the two bumps in the $M(|\phi_{\rm c}|)$ curve.
In the following, we label the three relevant branch types of
Fig.~\ref{fig:stability_schematic} as \textbf{S} (stable), \textbf{UM}
[unstable migrating, corresponding to the above scenario (iii)] and
\textbf{UB} [unstable BH, corresponding to scenario (iv)].

\subsection{Time evolutions of single boson stars}
\label{sec:timeevol}
%
\begin{figure*}
  \centering
  \includegraphics[width=0.86\textwidth]{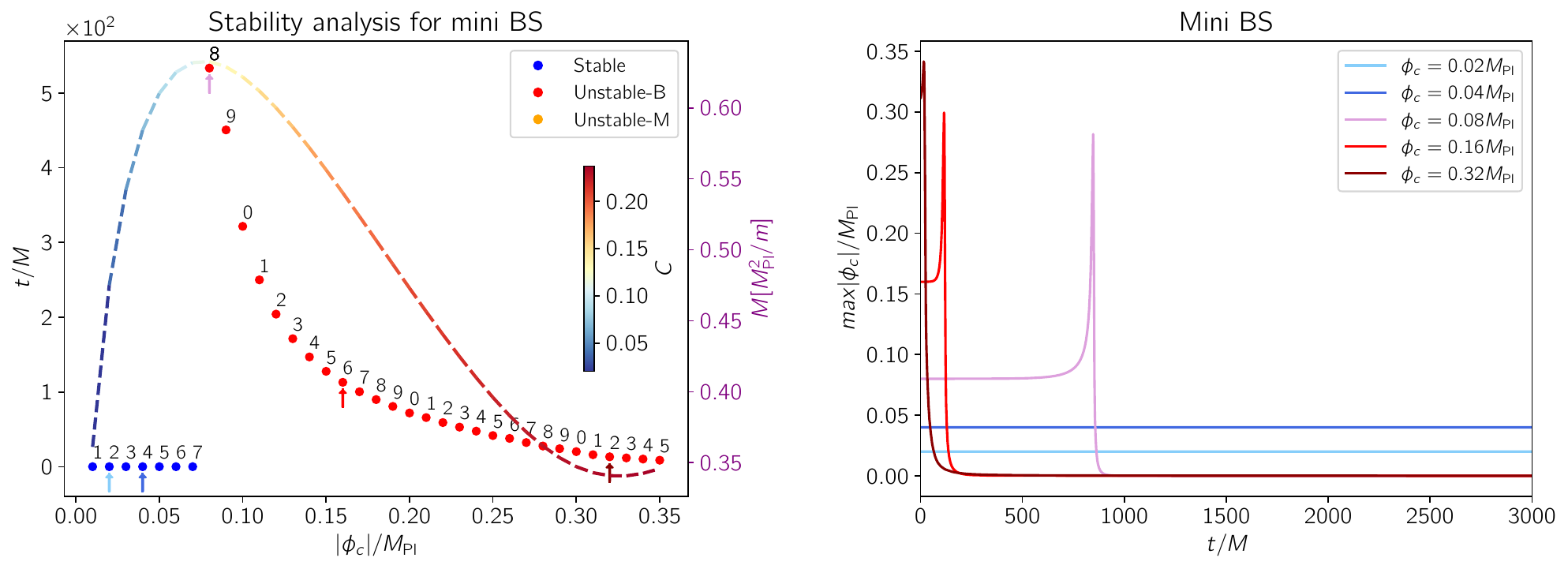}
  \\
  \includegraphics[width=0.86\textwidth]{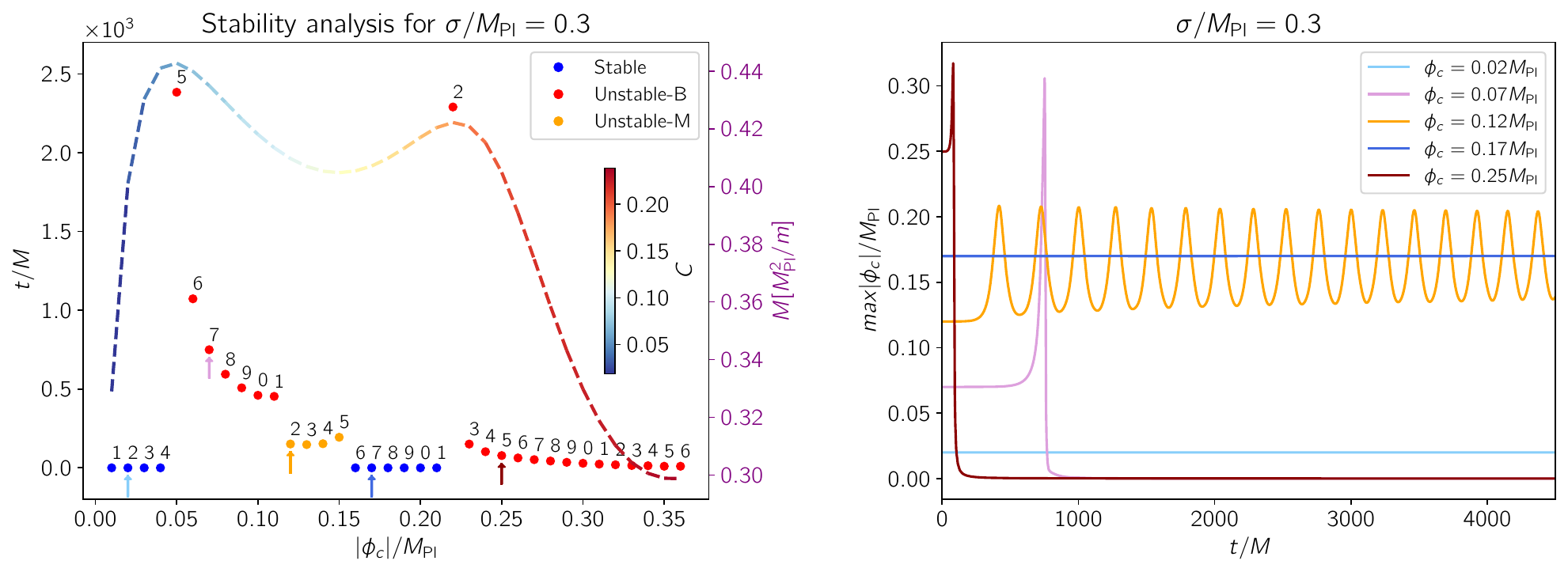}
  \\
  \includegraphics[width=0.86\textwidth]{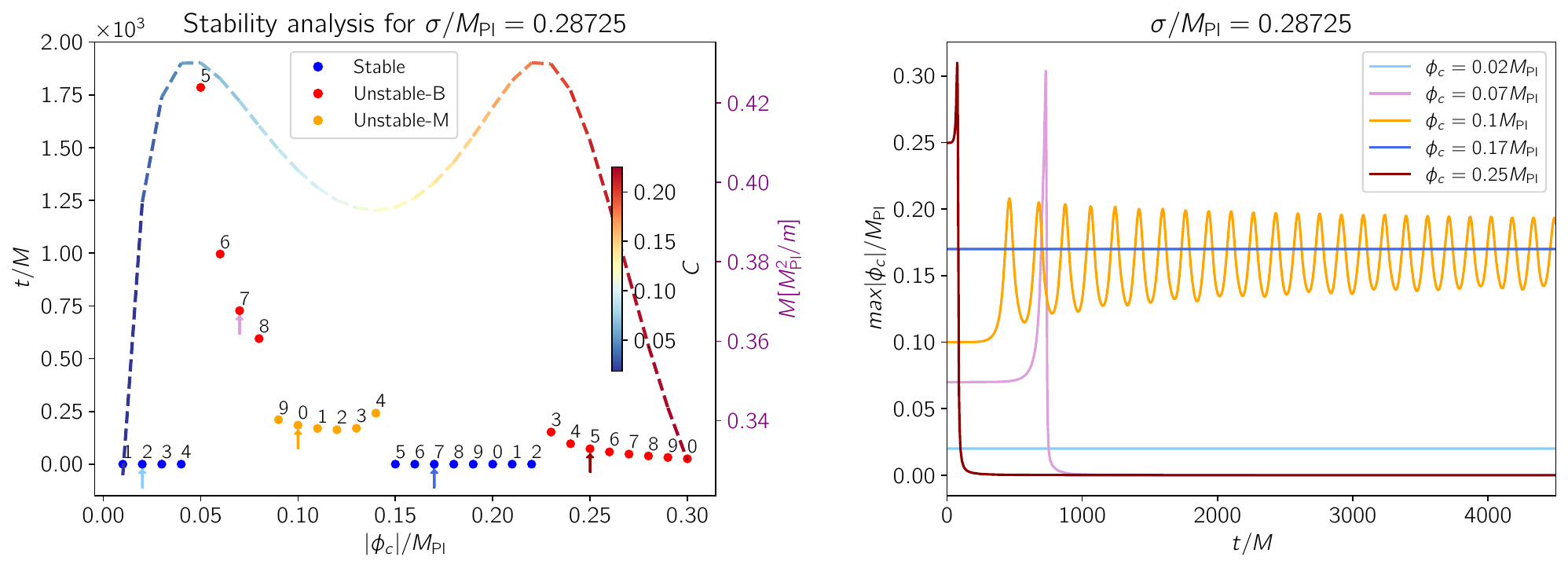} 
  \\
  \includegraphics[width=0.86\textwidth]{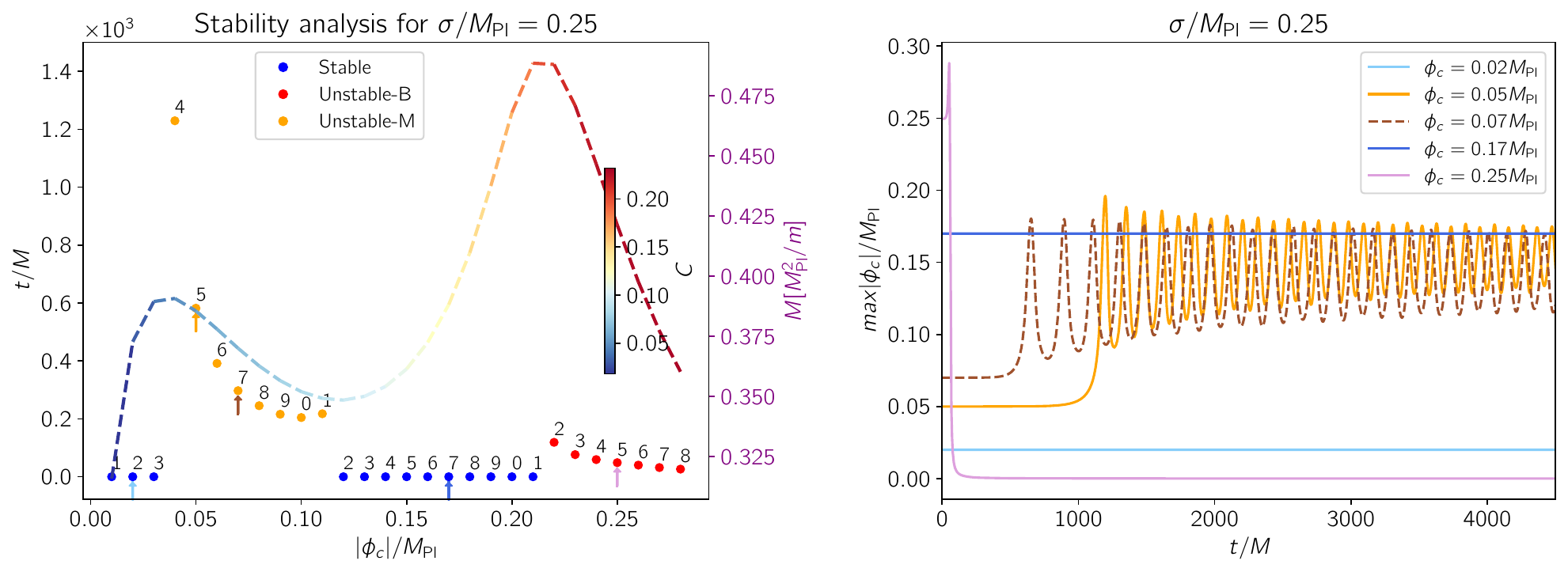}
  \caption{
  \textit{Left panels}: The $M(|\phi_{\rm c}|)$ curve, plotted for
  $\sigma/M_{\rm Pl} = \infty,~0.3,~0.28725,~0.25$, encodes in color
  the BS compactness $C$. The vertical axis on the left (right) marks the migration time (mass of the BS).
  The stability properties for selected
  stars are marked by colored symbols: blue for stable \textbf{S}
  branch stars, red for \textbf{UB} stars collapsing to a BH, orange
  for \textbf{UM} stars migrating to a more compact BS and their
  vertical position represents the migration time. The numbers
  adjacent to the symbols give the final digit of their $|\phi_{\rm
  c}|$ value, whilst the arrows underneath highlight the models
  illustrated in the corresponding right panels. 
  \textit{Right}: The time evolution of the scalar amplitude maximum,
  $|\phi|_{\rm max}(t)$.
  }
  \label{fig:dynamic_stability}
\end{figure*}
We investigate the dynamical behaviour of BSs from the three branch
types by evolving them in time with the numerical methods detailed
in Sec.~\ref{sec:numerics}. Here the instability is triggered by
imperfections in the initial data and/or discretization of the
evolution equations, but their particular nature is of significance
only in so far as they are of the same kind for all configurations.
In our analysis of these numerical  simulations,
we use the following diagnostics.
\begin{list}{\rm{\arabic{count}.}}{\usecounter{count}
             \labelwidth0.5cm \leftmargin0.7cm \labelsep0.2cm \rightmargin0cm
             \parsep0.5ex plus0.2ex minus0.1ex \itemsep0ex plus0.2ex}
\item
We monitor the maximum scalar-field amplitude $|\phi|_{\rm max}$
on the spatial domain as a function of time; we always find this
maximum to be realized at the BS center.
\item
We define the moment when $|\phi|_{\rm max}$ deviates from its
initial value by $0.1\,\%$ as the \textit{migration time}.
\item
We monitor the minimum value of the CCZ4 conformal factor function
$\chi$ on the spatial domain as a function of time. If $\chi_{\rm
min}$ drops below $0.1$, we conclude formation of a BH. We independently
verify this by checking that the scalar amplitude $|\phi|_{\rm max}$
exhibits a rapid drop to zero as the scalar field collapses into a
singular point that is not resolved numerically. We always find both BH
indicators in agreement.
\item
We quantify the compactness of equilibrium BSs as the maximal ratio
of the mass aspect to areal radius, $\displaystyle C\defeq \max_{r\ge
0} \tfrac{M(r)}{r}$.
\end{list}
The results of this analysis are summarized in
Fig.~\ref{fig:dynamic_stability}, which displays the evolution of
BSs across a wide range of compactness for the mini BS potential
$\sigma\rightarrow \infty$ and the solitonic cases $\sigma/M_{\rm
Pl} = 0.3$, $0.28725$ and $0.25$.  

Let us first consider the case
of mini BSs shown in the top row of the figure.  In complete agreement
with Fig.~\ref{fig:multistability}, the global maximum, i.e.~Kaup
limit, at $|\phi_{\rm c}|\approx 0.0765\,M_{\rm Pl}$ separates the
stable and unstable branches: all models to the left of this point
stay in their initial configuration (the blue symbols placed at
migration time $t=0$) whereas all models to the right collapse to
BHs (red symbols, marking \textbf{UB} instability). As expected,
the migration time away from their initial state exhibits a pole
at the Kaup limit, commensurate with the vanishing of the fundamental
radial eigenmode's frequency\footnote{We note Ref.~\cite{Santos:2024vdm}
reports that extremal-mass configurations do not necessarily
correspond to zero-frequency modes; our findings neither confirm nor
contradict their results.} \cite{Gleiser:1988ih}, and rapidly
drops with increasing compactness. In our simulations, models beyond
the mass minimum at $|\phi_{\rm c}| \approx 0.33$, despite being
on the potentially stable branch III of Fig.~\ref{fig:stability_schematic},
ubiquitously collapse to BHs as shown by the time evolutions
$|\phi|_{\rm max}(t)$ for selected BSs in the top right panel.

As we decrease $\sigma$ in the following rows of
Fig.~\ref{fig:dynamic_stability}, we see the second bump in the
$M(|\phi_{\rm c}|)$ curve moving towards smaller $|\phi_{\rm c}|$
and substantially increasing in magnitude. For all cases $\sigma
\le 0.3$, this second bump results in two sets of stable BSs --
cf.~the blue dots in the left panels of Fig.~\ref{fig:dynamic_stability}
-- corresponding to branches I and III of
Fig.~\ref{fig:stability_schematic}.  All models beyond the second
bump, i.e.~all models belonging to branch IV, are unstable and
ubiquitously collapse to BHs (red dots); for the cases
$\sigma\le 0.3$ in Fig.~\ref{fig:dynamic_stability}, this transition
to branch IV occurs at $|\phi_{\rm c}| \approx 0.22$.

The most complex behavior
arises for the moderately compact stars on branch II where $\du
M/\du |\phi_{\rm c}|<0$. For $\sigma=0.3$ and $\sigma=0.28725$,
this branch II hosts \textbf{UB} models collapsing to a BH on its
left end and \textbf{UM} models on its right end which migrate to
a more compact stable BS.  Evidently, this behavior depends on the
height of the second bump in the $M(|\phi_{\rm c}|)$ curve relative
to the first: initial configurations on branch II{\scriptsize UM},
i.e.~deep down in the trough, have relatively low mass and can
migrate horizontally\footnote{Strictly speaking, the migration is
not exactly horizontal in this diagram because the Noether charge,
rather than the BS mass, is conserved; the two quantities are
typically comparable in magnitude, however, so that the migration
occurs horizontally to leading order.  } at approximately fixed $M$
to the right where they find a stable configuration to settle down
to.  In the right panels of the figure, these stars exhibit a sudden
increase in the scalar $|\phi|_{\rm max}(t)$ followed by a slowly
damped oscillation around the target configuration's equilibrium
value.  BSs starting on branch II{\scriptsize UB} (further left),
in contrast, have too large a mass to find their collapse halted
by a stable BS configuration on the second bump and thus are doomed
to form a BH. In practice, we generically find this \textbf{UB}
segment of branch II to extend a bit further to the right than would
be expected from the height of the second bump; we believe this to
result from an overshooting of the dynamical compression of the
star beyond the compactness of its stable equilibrium cousin with
equal Noether charge. Finally, for the case $\sigma=0.25$, the second
bump has become so much larger relative to the first one that the
\textbf{UB} part of branch II has disappeared; every branch II star
now migrates to a stable BS configuration on branch III.

\section{BS collisions}
\label{sec:collisions}
%

\subsection{General behaviour}
\label{sect:general_behaviour}
The first and perhaps most elementary of our observations in
Fig.~\ref{fig:appetizer} is \emph{the enhanced radiative efficiency
of BS collisions as compared to their BH counterparts}. This has
already been noticed for oscillatons in Ref.~\cite{Helfer:2018vtq}
(see their Fig.~1) and in Ref.~\cite{Evstafyeva:2022bpr} for
unequal-mass BS collisions (see their Fig.~9). We interpret this
feature in terms of two competing effects: the GW emission is
expected to increase with (i) the binary constituents' compactness
and (ii) a higher degree of asymmetry around merger.  BSs, being
less compact but more vulnerable to deformation at merger than BHs,
are thus expected to have their radiative efficiency reduced by the
first effect but enhanced by the second.  Empirically, our results
confirm the observations of the above literature, namely that the
net effect is optimized for moderate compactness considerably below
that of BHs.
In this section, we will for now assert this empirical
result which we colloquially phrase as ``mergers of deformable BSs are
louder\footnote{Obviously, when the BS are \emph{too} squishy, 
they approach the weak field limit and GW production becomes less 
efficient.}'' in order to capture the fact that GW efficiency depends
on both, compactness and the tendency of BSs to deform into highly
asymmetrical configurations during merger.
We will then return to conjecture on
its origin in section \ref{sect:pot_dep_GW_E}.

Guided by the stability properties and corresponding migration or
collapse tracks of single BSs as summarized in
Fig.~\ref{fig:dynamic_stability}, we now aim to understand the
remaining phenomenology of the GW emission in head-on collisions
of BSs listed at the end of Sec.~\ref{sec:precis}: discontinuities
in the functional relation $E_{\rm GW}(|\phi_{\rm c}|)$; the
``needles'', i.e.~large and very sharp swings in the total GW energy
as one varies $|\phi_{\rm c}|$ for a fixed potential $V(|\phi|)$;
the correlation between $E_{\rm GW}(|\phi_{\rm c}|)$ and $M(|\phi_{\rm
c}|)$.
\begin{figure*}[hbt!]
  \centering
  \includegraphics[width=0.7\textwidth]{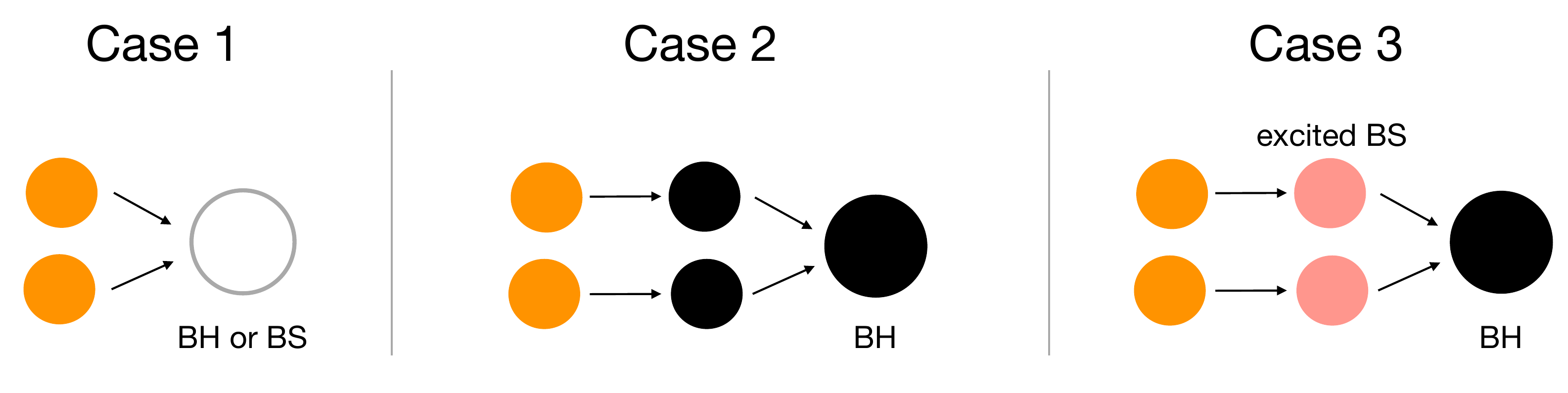}
  \caption{
  Illustration of 3 main fates of an equal-mass BS head-on merger
  encountered in this work. Case A corresponds to initially stable
  BSs either forming a BH or a BS remnant upon merger, case B to
  initially unstable BSs collapsing to BHs pre-merger, and case C
  to initially unstable BSs migrating to excited BS models and
  continuing their collapse to a BH after merger. We note that in
  case C we cannot exclude the possibility that a BS remnant is
  formed post-merger, however, in all of our simulations for this
  case a BH is formed.
  }
  \label{fig:merger_types}
\end{figure*}

For this purpose, we first summarize in Fig.~\ref{fig:merger_types}
the main types of mergers suggested by our stability analysis of
Fig.~\ref{fig:dynamic_stability}; there are three such types.  (1)
The initial BSs are stable, remain in equilibrium until merger where
they coalesce into a BS or BH remnant. (2) The initial BSs are
unstable, each of them collapses to a BH, and thenceforth they
complete their coalescence as a BH binary. (3) The initial BSs are
unstable but migrate towards a stable BS configuration around which
they oscillate in dynamic fashion; upon merging, they from a BH
remnant\footnote{Theoretically, they could also coalesce into a
more massive BS at merger, but we ubiquitously find them to form a
BH.}\,.

We next consider the repercussions of these three possible merger
scenarios in the context of our schematic stability Figure
\ref{fig:stability_schematic}. Starting from the left (small
$|\phi_{\rm c}|$), we encounter the following qualitative screenplay.
\\[5pt]
\noindent
\textit{\textbf{Branch I}}:
On branch I, we have stable BSs merging according to case 1 into a
BS or BH remnant. In consequence of the competing effects of
increasing compactness and decreasing susceptibility to deformation,
we expect the GW energy $E_{\rm GW}$ to rise as $|\phi_{\rm c}|$
is increased from 0, possibly peaking at some optimal compactness.
\\[5pt]
\noindent
\textit{\textbf{Branch II{\scriptsize UB} (red)}}:
Beyond point A, i.e.~the first maximum in the $M(|\phi_{\rm c}|)$
curve of Fig.~\ref{fig:stability_schematic}, we reach the red patch
of the unstable branch II; our initial BSs are unstable, collapse
to individual BHs during infall and form a BH merger remnant at
coalescence according to the case 2 scenario of
Fig.~\ref{fig:merger_types}. The radiated energy function $E_{\rm
GW}(|\phi_{\rm c}|)$ should consequently exhibit a discontinuous
jump from a large BS value to the generic $E_{\rm GW}^{\rm BH}\approx
6\times 10^{-4}\,M_{\rm tot}$, as has already been observed for
head-on collisions of oscillatons
\cite{Helfer:2018vtq}.

Let us next connect these theoretical expectations with the results
from the actual BS collisions displayed in Fig.~\ref{fig:appetizer}.
There we indeed observe the expected increase in $E_{\rm GW}(|\phi_{\rm
c}|)$ at low compactness. The curve furthermore exhibits a local
maximum before rapidly dropping to $E_{\rm GW}^{\rm BH} \approx
6\times 10^{-4}\,M_{\rm tot}$ marked by the red dots
around $0.068\lesssim |\phi_{\rm c}|/M_{\rm Pl} \lesssim 0.083$.
Contrary to our expectation, however, this rapid drop occurs at
$|\phi_{\rm c}|$ values {\it beyond} the maximum mass (cf.~the
dashed purple curve in the figure). As we will discuss in more
detail in Sec.~\ref{sect:initial_sep}, this discrepancy between
our results and theoretical expectations is a consequence
of the finite infall time which, for marginally unstable BSs,
is insufficient for the stars to complete their collapse to a
BH; binaries marginally to the right of transition point A therefore
coalesce as BSs of increasing compactness but not yet as BHs.
This effect also introduces a continuity of $E_{\rm GW}(\phicen)$
contrary to the discontinuity expected in the limit of infinite
infall time.
\\[5pt]
\noindent
\textit{\textbf{Branch II{\scriptsize UM} (yellow)}}:
Beyond transition point B, on branch II{\scriptsize UM} in
Fig.~\ref{fig:stability_schematic}, we encounter unstable stars
which migrate horizontally to other stable and more compact BS
models and eventually merge according to case 3.  The transition
at B is discontinuous in the sense that the compactness of these
equilibrium target configurations differs from that of a BH by a
finite amount. We accordingly expect an upward jump in $E_{\rm
GW}(|\phi_{\rm c}|)$. This expectation is fully corroborated in
Fig.~\ref{fig:appetizer}; at $|\phi_{\rm c}|=0.083\,M_{\rm Pl}$,
the GW energy \emph{discontinuously} increases by more than a factor
two and a BH only forms post-coalescence (red versus blue dots).

Branch II{\scriptsize UM} is also the home of the ``needles'',
i.e.~$\mathcal{O}(1)$ relative changes in $E_{\rm GW}$ under miniscule
variations in $|\phi_{\rm c}|$.  We can explain this behavior by
recalling the single migrating BS models of
Fig.~\ref{fig:dynamic_stability}. These stars do not complete their
migration by quickly settling down into the new equilibrium
configuration but rather pulsate around it; cf.~the orange curves
for $|\phi_{\rm c}(t)|$ in the right panels. In a head-on collision
of two such stars, this pulsation is abruptly terminated by
the merger process. The merger dynamics and consequential GW energy,
however, will depend on the momentary phase of the pulsation; do
the two stars merge at the moment of maximal expansion or contraction
or somewhere in between? It is this pulsation phase and the
consequential BS compactness {\it at merger} that changes by
$\mathcal{O}(1)$ even under a minute change in the initial BSs'
central scalar-field amplitude value $|\phi_{\rm c}|$. The energy
$E_{\rm GW}$, as we have already seen, changes accordingly with the
BS compactness at merger.

Clearly, since the phase at merger depends on the time of merger,
we should expect the structure of the needles to change when we
change the initial separation (and hence the infall time). As we
will see later in section \ref{sect:initial_sep}, this is indeed
what occurs when we vary the initial separation.

\noindent
\textit{\textbf{Branch III}}:
The transition at point C, where we reach the stable stars of branch
III, is continuous in every regard. And yet, point C is special in
one important respect: it marks the compactness where the energy
function $E_{\rm GW}(|\phi_{\rm c}|)$ must be {\it extremal}. This
can be understood as follows. The branch II{\scriptsize UM} stars
located an infinitesimal $\delta|\phi_{\rm c}|$ to the left of point
C are unstable but migrate horizontally to a new equilibrium
configuration also infinitesimally close to point C, just on the
right. In consequence, the merger scenarios and, hence, the GW energy
values $E_{\rm GW}$ of these two slightly different initial
configurations must be {\it identical}; it does not matter whether
the stable BS states have been reached through a minor migration
or were realized right from the beginning. Thus, $E_{\rm GW}$
infinitesimally to the left of point C must equal the value
infinitesimally to the right of point C and, hence, must be extremal
and smooth at C. Up to minor distortions due to the presence
of the needles, we see this behaviour confirmed in Fig.~\ref{fig:appetizer}
at $|\phi_{\rm c}|\approx 0.14$. As we shall see further below,
this remains true for other potential functions {\it provided}
transition point C exists and admits stable BS models in its
neighbourhood.
\\[5pt]
\noindent
\textit{\textbf{Branch IV}}:
The transition to branch IV at point D is very similar to that of
point A between the stable and unstable branches I and II{\scriptsize
UB}.  Theoretically we expect the same \emph{discontinuous} jump
in the radiated energy back to the BH value $E_{\rm GW}^{\rm BH}
\approx 6\times 10^{-4}\,M_{\rm tot}$ and the associated premature
BH collapse beyond point D. Unlike point A, our empirical results
fully reproduce the theoretical expectations in Fig.~\ref{fig:appetizer}
at $|\phi_{\rm c}|\approx 0.22\,M_{\rm Pl}$. This excellent agreement
directly follows from the generically high BS compactness in this
regime; all unstable BSs collapse to BHs rapidly, well within the
infall time. Note also that for these configurations there exists
no stable BS branch further to the right which might halt or slow
down the collapse. \\[5pt]
%

\subsection{GW Energy dependence on compactness}
\label{sect:pot_dep_GW_E}

We now return to the origin of the empirical result that mergers
of less compact BS stars are significantly louder than equivalent
black hole mergers. To understand this, let us consider the dependence
of the GW energy $E_{\rm GW}(|\phi_{\rm c}|)$ as a function of
$\phicen$, first focusing on the $\sigma/\mpl= 0.28725$ models
and then generalizing to other potentials.

Starting from the small $\phicen$ limit of Branch I, as $\phicen\rightarrow
0$,  we approach the trivial minimum $V(\phicen \rightarrow 0) =0$,
so the BS becomes more and more diffuse. Thus we expect in this
limit, the GW energy also approaches zero which is manifest in
Fig.~\ref{fig:appetizer}.  As we increase $\phicen$ from this limit
and climb the potential, the BS compactness increases, so the GW energy
$E_{\rm GW}(|\phi_{\rm c}|)$ also increases. This is an unsurprising
consequence of the usual expectation that the GW energy is positively
correlated with the compactness. What \emph{is} surprising however
is that by $\phicen=0.03\mpl$, $E_{\rm GW}(|\phi_{\rm c}|)/M_{\rm
int} \approx 0.001$ \emph{which is already greater than that of an
equivalent BH head one merger with $E_{\rm GW}^{\rm BH} = 0.0006$},
despite the BSs being significantly less compact.

Obviously, this  positive correlation cannot continue indefinitely
to higher compactness since eventually one reaches BH compactness.
Focusing on Fig.~\ref{fig:appetizer}, and ignoring for now the messy
needles and instabilities that dominate the physics for mid values
of $\phicen$ on Branch II, we see that indeed there exist a turnover
at $\phicen\gtrsim 0.13\mpl$ (roughly from the beginning of the stable
Branch III) when the transition from positive correlation to negative
correlation occurs -- $E_{\rm GW}(|\phi_{\rm c}|)$ begins to decrease
even as the compactness increases. Finally at $\phicen = 0.22\mpl$,
the GW power discontinuously jumps to the expected GW emission
$E_{\rm GW}^{\rm BH}=0.0006$ from BH collisions -- this signals the
onset of the case 2 mergers (see Fig.~\ref{fig:merger_types}) since
$\phicen=0.22\mpl$ is a critical point of BS instability.

Thus we have two seemingly paradoxical results: (a) why should BSs
that are less compact than a BH generate more GWs in a merger and
(b) why do BSs of very high compactness generate lower GW energy
than BSs of low compactness (but still more than BHs)?

For question (a), we have argued in
Sec.~\ref{sect:general_behaviour} that in a BS merger a greater
asymmetry of the coalescence phase generates stronger gravitational
radiation. Observation (b) can then be explained by the argument
that \emph{higher compactness implies lower deformability}. What
we mean by the vague term ``deformability'' is ``the amplitude of
the response of the BS system to gravitational and matter perturbations''
as quantified through {\it tidal deformability parameters}; see
e.g.~Ref.~\cite{Sennett:2017etc}. Since most of the GW emission in a
head-on merger is generated during the coalescence phase, it follows
that the larger the initial asymmetry of the apparent horizon or
compact matter remnant, the higher the GW emission
during the merger and ringdown would be. A high compactness, on
the other hand, implies a deeper gravitational potential well where
matter fields are less susceptible to perturbations during a merger
-- they are ``stiffer''. Clearly, this effect is competing with the
fact that systems need to be strongly gravitating before GWs can
be produced since spacetime itself is also extremely stiff, hence
there is a maximum -- in the case of $\sigma/\mpl = 0.28725$, this
occurs at\footnote{It has been argued in Ref.~\cite{Okounkova:2020vwu}
that in the strong-field regime, non-linearities around merger are
rapidly encompassed by a common horizon (see their Sec.~II B) and
thus will not reach a detector outside the event horizon. Our
observation of enhanced GW energy from more asymmetric mergers is
independent of this interpretation of captured non-linearities.}
$\phicen \approx 0.13$ where $E_{\rm GW}(|\phi_{\rm c}|)/M_{\rm
tot} =0.0035$.

Finally, we can infer the point of maximal GW emission by recalling
that $E_{\rm GW}(\phicen)$ is extremal at
transition point C connecting branches II and III, provided
point C admits stable models in its neighbourhood.
Ignoring for the moment the needles and troughs
due to premature BH collapse in $E_{\rm GW}(\phicen)$, we can furthermore argue
as follows that the extremum is likely to
be a maximum: First, we know that $E_{\rm GW}(\phicen)$ is a rising
function at very small $\phicen$. $E_{\rm GW}$ can then be minimal at
point C only if it has a maximum somewhere between $\phicen=0$ and
point C. To the left of point C, we would therefore already be in the
regime where increasing compactness lowers GW emission. It is
hard then to conceive of a physical mechanism that would lead to
yet another reversal of $E_{\rm GW}(\phicen)$ towards a rising slope.
We cannot rigorously rule out that some such mechanism exists, but
parsimony suggests the transition point
C simply marks the transition from increasing $E_{\rm GW}$ (due to
increasing compactness) towards decreasing $E_{\rm GW}$ (due to
reduced asymmetry). Empirically, we find this view fully confirmed;
if point C exists and admits stable BSs in its neighbourhood,
it marks {\it maximal} GW emission.


\begin{figure*}
  \centering
  \includegraphics[width=0.48\textwidth]{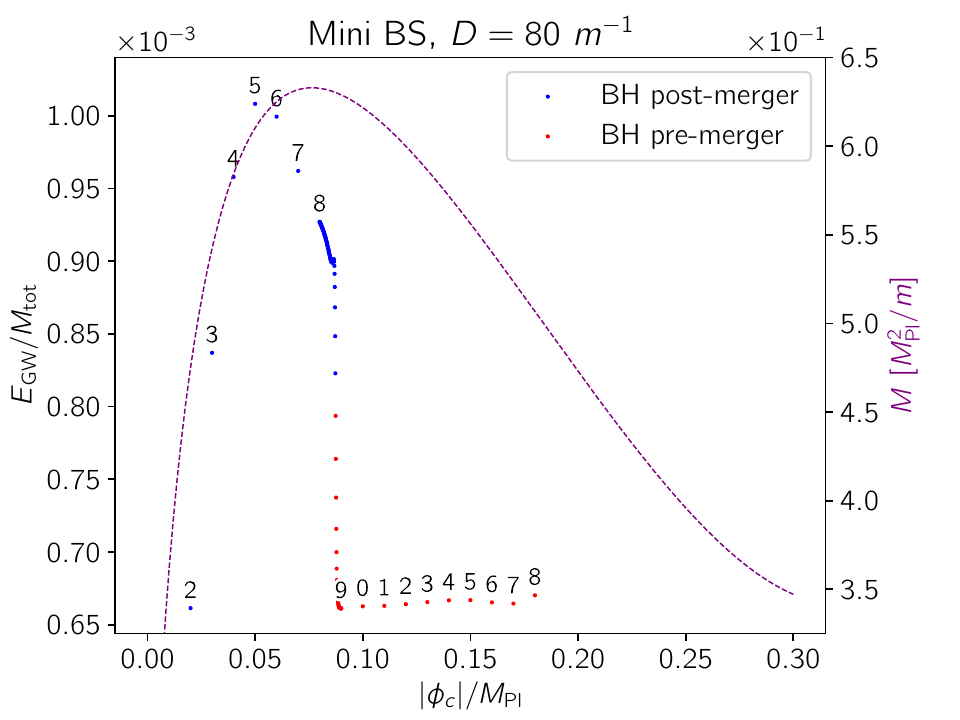}
  \includegraphics[width=0.48\textwidth]{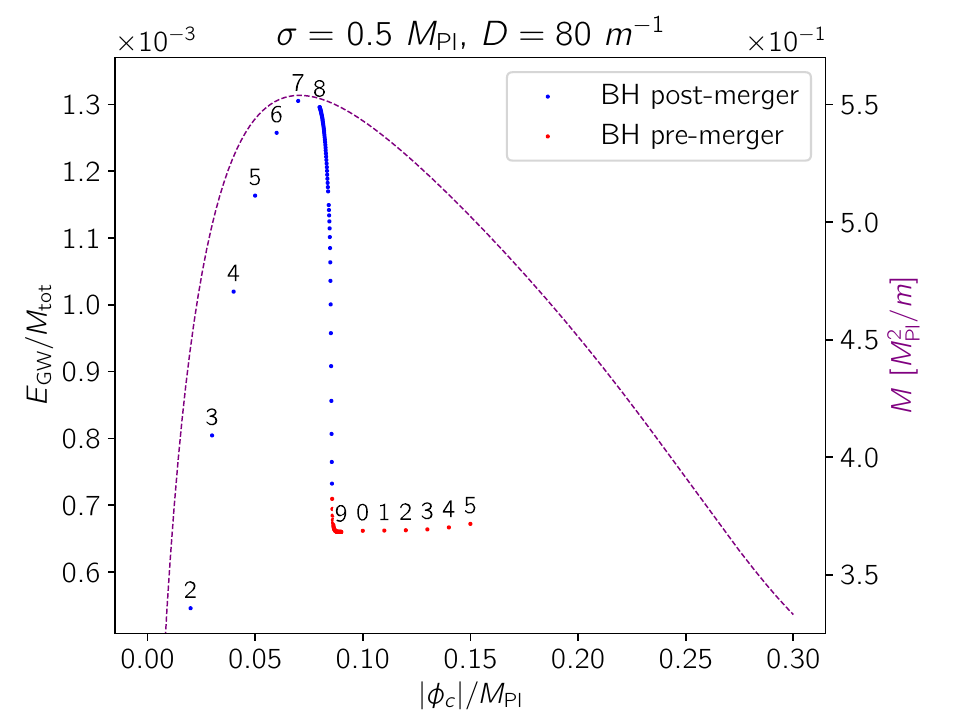}
  \includegraphics[width=0.48\textwidth]{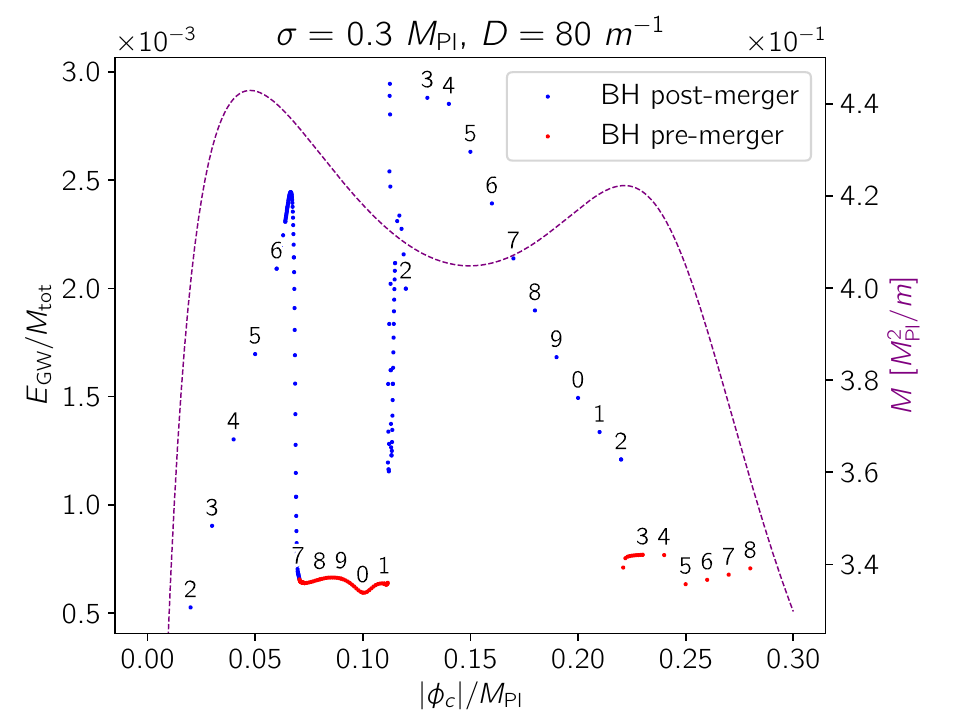}
  \includegraphics[width=0.48\textwidth]{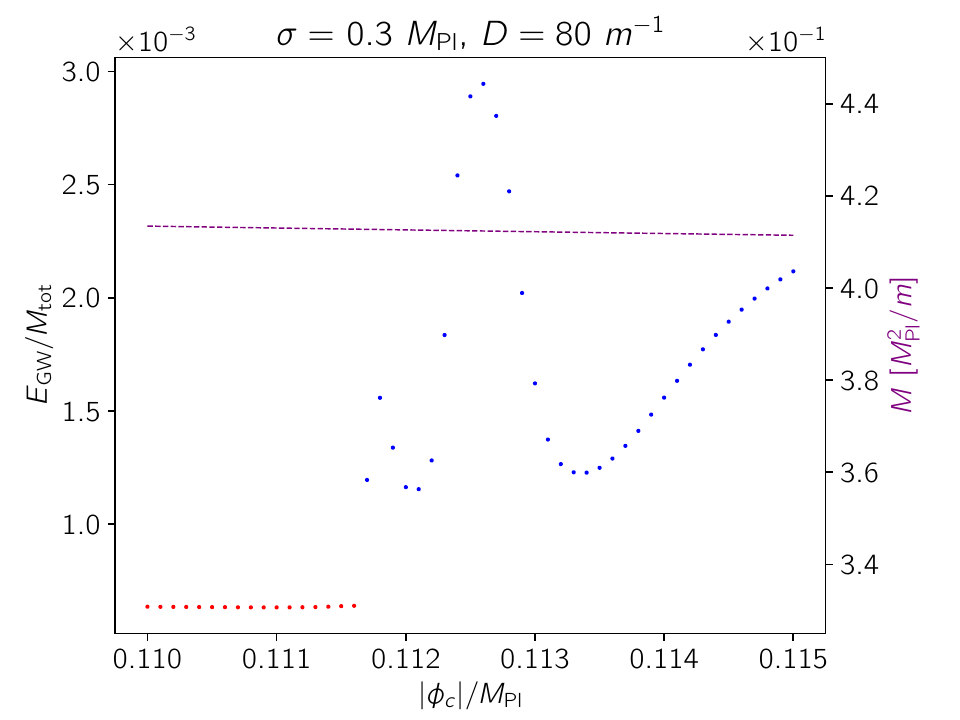}
  \includegraphics[width=0.48\textwidth]{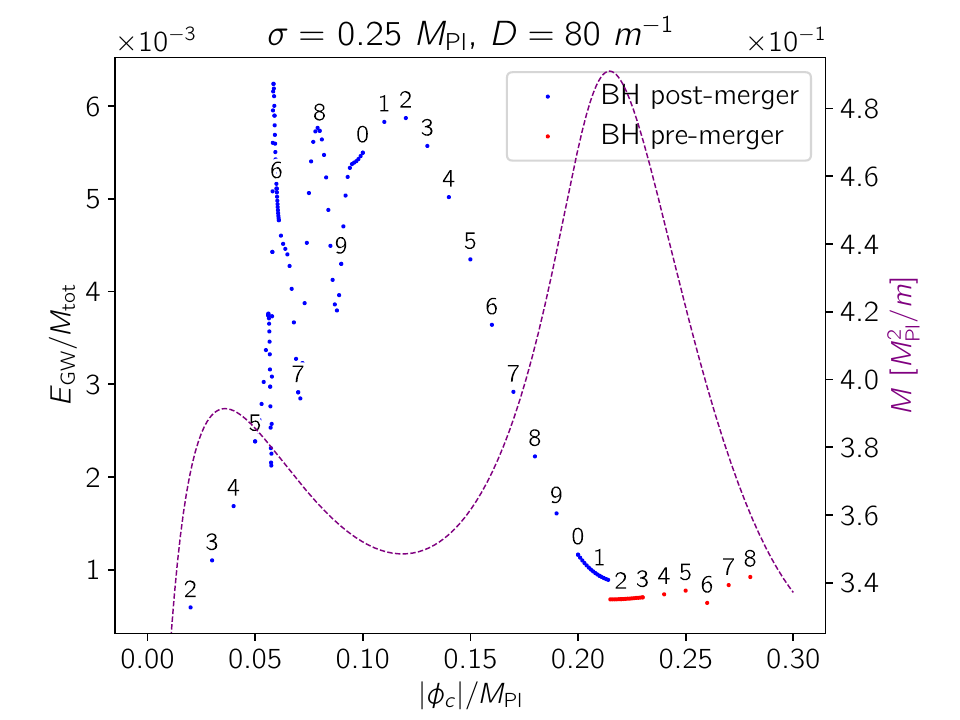}
  \includegraphics[width=0.48\textwidth]{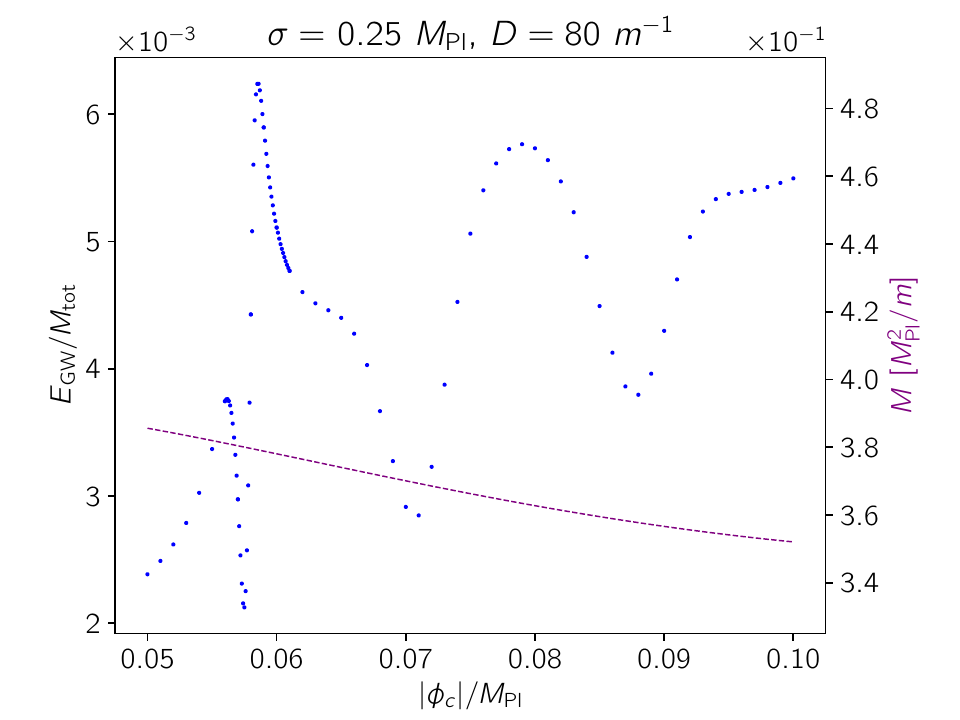}
  \caption{
  As in Fig.~\ref{fig:appetizer} but for different scalar potentials
  $V(|\phi_{\rm c}|)$, this figure shows the GW energy released in
  equal-mass BS head-on collisions as a function of the individual
  BSs' central scalar-field amplitude $|\phi_{\rm c}|$.  Here, red
  dots denote configurations where the
  individual BSs collapse into BHs prior to coalescence and blue
  symbols represent binaries where the
  constituents retain their BS character up to merger upon which
  they form a single BH or BS remnant. For reference, we also display in
  each panel by the dashed purple curve the $M(|\phi_{\rm c}|)$,
  i.e.~mass vs.~scalar amplitude, curve of the corresponding
  equilibrium BS models. Note that the center- and bottom-right
  panels are zoom-in versions of the respective left panels.
  }
  \label{fig:EGW}
\end{figure*}


\subsection{GW energy dependence on $V(\phi)$}
Up to this point, we have discussed for the solitonic potential
with $\sigma=0.28725\,M_{\rm Pl}$ of Fig.~\ref{fig:appetizer} all
possible features that can arise in the GW emission of BS head-on
collisions of different compactness and we have related these
features to the types of merger sketched in Fig.~\ref{fig:merger_types}.
The merger cases 1 and 2 will be realized for any potential that
admits stable and unstable BSs, but the presence or absence of case
3 will critically depend on the relative location of the two maxima
in the $M(|\phi_{\rm c}|)$ curve labelled A and D in
Fig.~\ref{fig:stability_schematic}. Notice from Fig.~\ref{fig:multistability}
that both the position
and amplitude of the 2nd peak D depend on $\sigma$ -- the larger
$\sigma$, the larger the $\phicen$ value of transition point D is
while the amplitude of its potential $V(\phicen)$ becomes smaller.
Thus depending on the amplitude of $\sigma$, we expect three
different possibilities.
\begin{list}{\rm{(\roman{count})}}{\usecounter{count}
             \labelwidth0.5cm \leftmargin0.7cm \labelsep0.2cm \rightmargin0cm
             \parsep0.5ex plus0.2ex minus0.1ex \itemsep0ex plus0.2ex}
\item
In the mini BS limit where $\sigma \rightarrow \infty$, the second
peak D occurs at very large compactness, i.e.~$|\phi_{\rm c}|$
such that all branch III stars are
unstable. In this scenario, branches II and III effectively merge
with branch IV into one single unstable \textbf{UB} branch and the GW energy
$E_{\rm GW}(|\phi_{\rm c}|)$ displays a simple behaviour, radiatively
efficient BS mergers for small $|\phi_{\rm c}|$ and $E_{\rm
GW}^{\rm BH}\approx 6\times 10^{-4}\,M_{\rm tot}$ for large
$|\phi_{\rm c}|$. This is exactly the behaviour we observe in the
top row of Fig.~\ref{fig:EGW} for mini BSs and $\sigma=0.5\,M_{\rm
Pl}$.  \item  On the other hand, for small $\sigma/M_{\rm Pl}
\lesssim 0.275 $,  the second peak D occurs at sufficiently
small $|\phi_{\rm c}|$ such that branch III admits stable BSs and
the BS mass at point D (significantly) exceeds that of point A.
In that case, branch II{\scriptsize UB} is absent. Less compact
binaries, i.e.~starting to the left of point D, merge according
to types 1 or 3 with large $E_{\rm GW}$ while merger case 2 is
encountered for and only for highly compact binaries starting to
the right of point D. In Fig.~\ref{fig:EGW}, we encounter this
behaviour in the bottom row for $\sigma=0.25\,M_{\rm Pl}$.
\item
For moderate values $0.275\lesssim \sigma/M_{\rm Pl}\lesssim 0.4$,
on the other hand, the global maximum of $M(|\phi_{\rm c}|)$ is
realized at point A (``the first bump is higher'') and then we
encounter the entire program of $E_{\rm GW}(|\phi_{\rm c}|)$ as
summarized in the previous subsection. This is the case displayed
in Fig.~\ref{fig:appetizer} for $\sigma=0.28725\,M_{\rm Pl}$
and in the center row of Fig.~\ref{fig:EGW} for $\sigma=0.3\,M_{\rm Pl}$.
\end{list}
Results for further values of the solitonic potential parameter
$\sigma$ are presented in Ref.~\cite{Ge2024}; all of these fall into
one of the three regimes listed here.

\begin{figure*}[t]
  \includegraphics[width=0.95\textwidth]{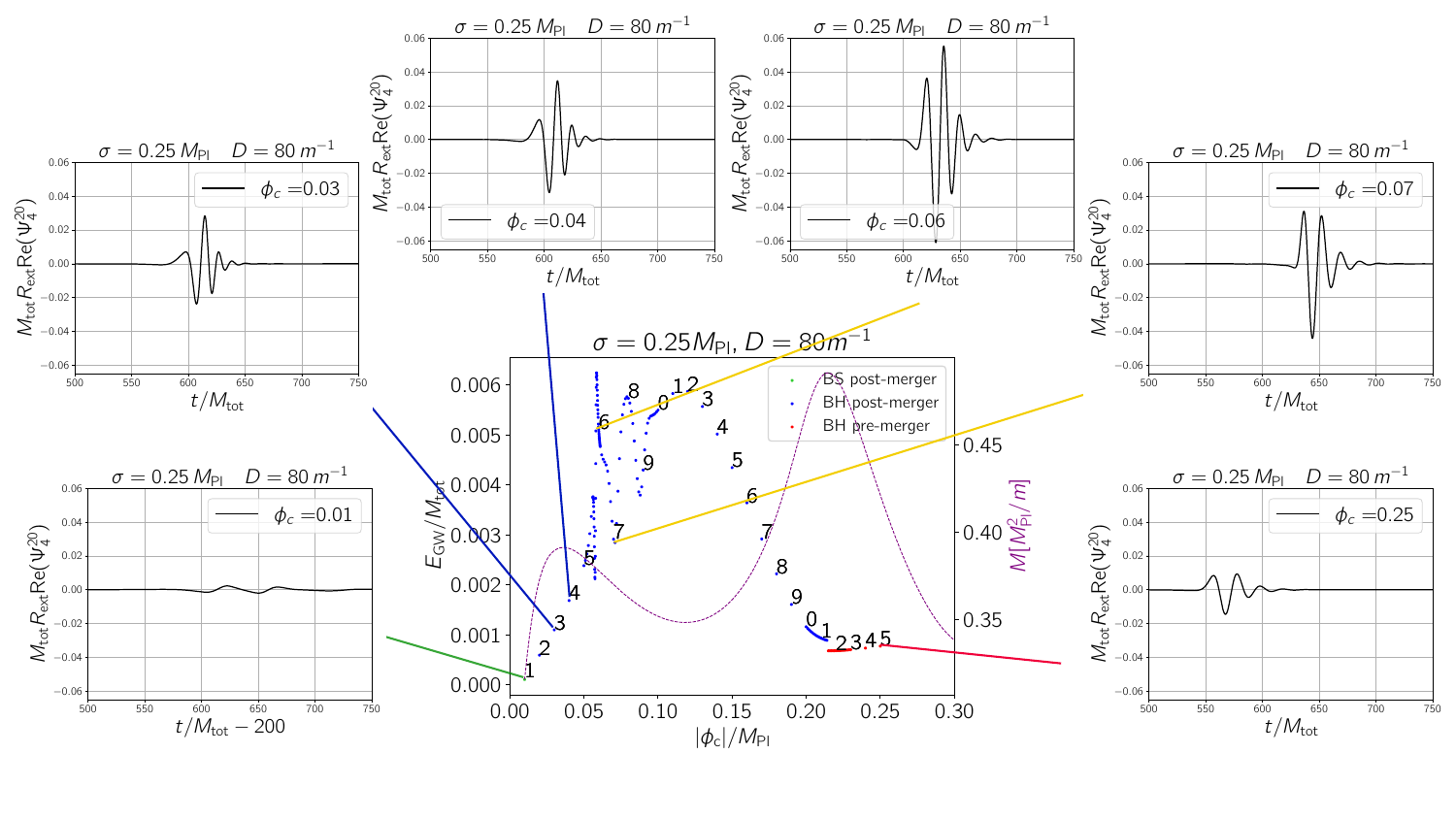}
  \caption{
  The GW signals from selected BS collisions for $\sigma=0.25 M_{\rm{Pl}}$ are
  shown in the form of the dominant quadrupole contribution of the
  Newman-Penrose scalar $\Psi_4$. The straight lines, color coded
  according to the branch colors of Fig.~\ref{fig:stability_schematic},
  connect the individual waveforms to the corresponding configuration in the
  $E_{\rm GW}(\phicen)$ curve. The signals exhibit a wide range of amplitudes,
  in accordance with the GW energy values of the central panel, but
  all agree quite closely in their overall shape. Note that for the  green line ($\phicen=0.01\mpl$ merger) is that of Case 1 (merger into another BS).
  }
  \label{fig:psi4_diagram}
\end{figure*}

The remarkably rich structure that we have encountered in the dependence
of the GW energy on the BS parameters is not accompanied by a corresponding
complexity in the GW signals. Rather, we universally observe rather
standard head-on waveforms with a shape quite indistinguishable by
eye from their BH counterparts. For example, we display in
Fig.~\ref{fig:psi4_diagram} the leading quadrupole mode of the
Newman-Penrose scalar $\Psi_4$ for $\sigma=0.25\,M_{\rm Pl}$ and
various scalar-field amplitudes including configurations of the
needle regime. All signals have the overall structure, also seen
for BHs, as for example in Fig.~8 of Ref.~\cite{Sperhake:2006cy},
and only differ significantly in their overall amplitude.

%

\subsection{Impact of the initial BS separation}
\label{sect:initial_sep}
In our discussion in section \ref{sect:general_behaviour}, we have
identified two features in the shape of the GW energy $E_{\rm
GW}(|\phi_{\rm c}|)$ which we expect to vary with the initial
separation of the binary. The first observation concerns 
BS mergers with
correspondingly large GW emission on the {\it early} part of branch
II{\scriptsize UB} to the right of transition point A, despite the
fact that they are unstable to BH collapse. How can this be?
The answer
lies in the inevitably finite infall time of our BS collisions; in
Fig.~\ref{fig:dynamic_stability}, the left, third from top panel
demonstrates that for $\sigma=0.28725$ and $|\phi_{\rm c}|\lesssim
0.06\,M_{\rm Pl}$, BSs are indeed unstable to BH collapse but take
a long time to do so, longer, in fact, than the duration of their
infall. They consequently merge as increasingly compact BSs but not
yet as BHs. At $|\phi_{\rm c}|=0.068\,M_{\rm Pl}$ we finally reach
BSs so unstable that they complete their individual collapse prior
to encountering each other.  This behaviour also explains the
continuity of the curve $E_{\rm GW}(|\phi_{\rm c}|)$ curve at $|\phi_{\rm
c}|=0.068\,M_{\rm Pl}$ in Fig.~\ref{fig:appetizer}; for each finite infall time, we can fine
tune $|\phi_{\rm c}|$ to obtain any ``desired'' BS compactness at
merger -- note that these BSs are {\it not} in equilibrium and,
hence, not constrained by the maximum compactness of equilibrium
stars.
We have tested this hypothesis by rerunning
selected configurations in this regime with different initial
separation $D$ but all other parameters unchanged. We indeed find in
these simulations that the sharp drop in the GW energy towards
$E_{\rm GW}^{\rm BH}$ moves towards larger $\phicen$ when we
decrease $D$ whereas it converges to the transition point A
in the limit $D\rightarrow \infty$.

\begin{figure}[tb]
    \includegraphics[width=0.48\textwidth]{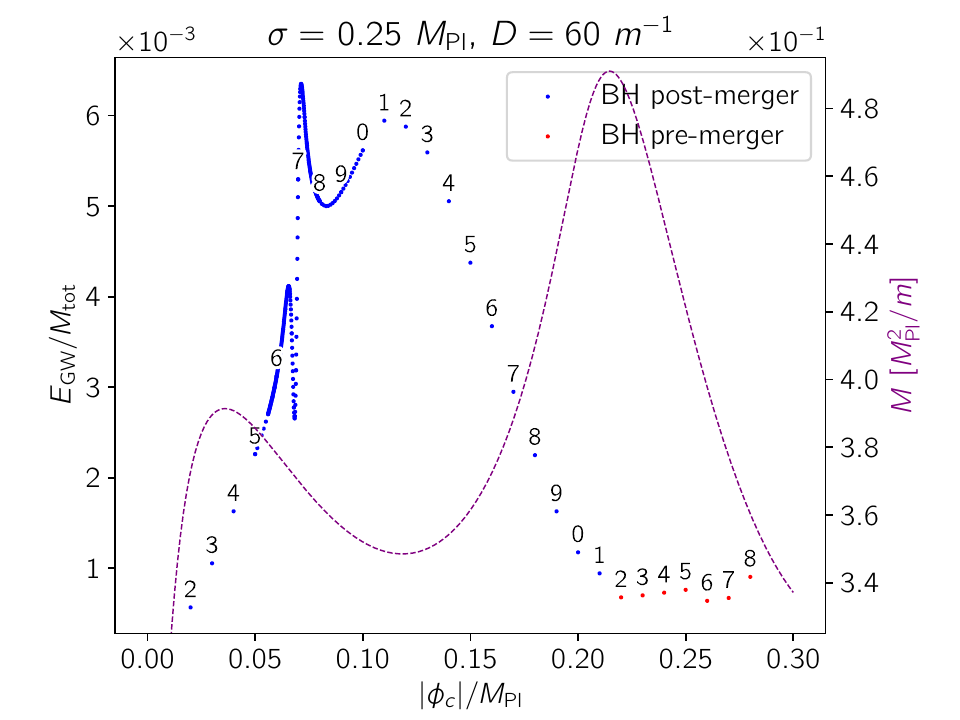}
  \caption{As Fig.~\ref{fig:EGW} but for $\sigma=0.25\,M_{\rm Pl}$
  and a smaller initial binary separation $D=60\,m^{-1}$.  Compared
  to the bottom left panel of Fig.~\ref{fig:EGW}, we see that the
  different initial separation significantly changes the structure
  of the needles in $E_{\rm GW}(|\phi_{\rm c}|$ around $|\phi_{\rm
  c}|=0.07\,M_{\rm Pl}$ but leaves the curve virtually unaltered
  everywhere else. 
  }
  \label{fig:sigma=0.25d60}
\end{figure}

The second feature is the structure of the needles under
variations in the initial separation; here a different
infall time implies a different
cut-off of the migrating BSs' radial pulsations. The structure
of the needles in the $E_{\rm GW}(|\phi_{\rm c}|)$ curve should
then change when varying $D$. We verify this interpretation
in Fig.~\ref{fig:sigma=0.25d60} where we plot the GW energy for
$\sigma=0.25\,M_{\rm Pl}$ but now with initial binary separation
$D=60\,m^{-1}$. Our expectation is fully borne out by comparing
this figure with the analogous results for $D=80\,m^{-1}$ in the
bottom left panel of Fig.~\ref{fig:EGW}; the needles have drastically
different shape while $E_{\rm GW}^{\rm BH}$ retains its overall
shape in every other regard.

\section{Discussion}
\label{sec:discussion}
In this work, we have undertaken an exhaustive numerical and theoretical
investigation of the physics of GW production in head-on boson stars
mergers, scanning over both the parameter space $\sigma$ of the
solitonic potential and compactness parametrized by the central scalar
amplitude $\phicen$ for
equilibrium BS solutions. While we have summarised the results in
the introduction section \ref{sec:intro}, it is worth asking which
of the results discussed are model specific to the solitonic model
we have studied and which are general broad lessons applicable to other
models.

The most striking result is the fact that BS mergers at high
compactness generate more gravitational radiation
than equivalent mergers of BHs.
This was first noted in studies of solitonic models in
Refs.~\cite{Helfer:2021brt,Evstafyeva:2022bpr} and in 
Ref.~\cite{Helfer:2018vtq} for the $m^2\phi^2$ model
of a real scalar field (sometimes called an \emph{oscillaton}). The
fact that it appears in both real and complex scalar models
suggests that \emph{head-on mergers of compact bosonic self-gravitating
stars generate significantly more GW energy than equivalent BH mergers}.
It is thus an interesting question to ask whether this result extends
to inspiral mergers of BSs.  Indeed, our results demonstrate that,
above some model-dependent compactness, BS mergers
are {\it always} louder than BHs, even if they eventually approach
(but do not reach) BH compactness -- ``deformable BSs are 
louder than stiff BHs''.
Our conjecture for this result -- supported
by the numerical observations of ``needles'' in the GW power during
mergers of unstable Branch II{\scriptsize UM} mergers -- is that
during the collision phase, the increased deformation in a less
compact BS results in a more strongly perturbed and asymmetric 
merger remnant whose
subsequent relaxation generates more GWs. {\it It will be interesting to
see if this result extends to other types of horizonless compact
objects such as Proca stars or perhaps even NSs.}

In general, BS models become unstable at high (but below BH) compactness
where perturbations can trigger a rapid collapse to a BH.
In both mini-BS and oscillaton models, this represents
the only delineation point between stable and unstable BSs. However,
in more complicated models with two or more minima in the potential,
the presence of additional stable and unstable equilibrium points
in the potential $V(\phicen)$ can introduce additional structure in the stability
diagram (see e.g.~Ref.~\cite{Helfer:2016ljl}). For the solitonic
models we have studied, this manifests itself in the form of additional
stable/unstable branches in the solution space of BSs (see Fig.~\ref{fig:multistability}). In these additional unstable regions, a
BS may either collapse to a BH or migrate to a different BS. In all cases,
the existence of these unstable regions implies that, in the
limit of infinite infall time, \emph{the GW
energy $E_{\rm GW}(\phicen)$ is discontinuous at the points
of transition from stable to unstable branches}. Unfortunately, since we do not
measure GW energies over extraction spheres directly, but only the
strain at one point on the sphere, there is scant hope for
observing such discontinuities.

In section \ref{sec:single}, we have studied in detail the nature
of the instability, in particular the time-scale or \emph{migration
time} of the transition.
Furthermore, if
the instability leads to a transition into another (excited) BS
which oscillates, then we should expect ``needles'' structure
in the energy function $E_{\rm GW}(\phicen)$.
The main lesson here is that, \emph{there can be
rapid changes in GW energy over small changes in the parameter
space}. For example, if we merge two equal mass oscillating BSs, the
GW emitted would depend strongly on the \emph{phase} of the pulsation
at merger,
complicating any observational inference over its progenitors.
This raises two further important questions. (i) {\it What is the timescale
of relaxation of dynamic BS configurations to an equilibrium state?}
Numerical studies often result in long-lived dynamical stars
(see e.g.~Refs.~\cite{Croft:2022bxq,Siemonsen:2023hko} or
our Fig.~\ref{fig:stability_schematic}), but is it long
enough to be of relevance for inspirals? (ii) {\it What kind of
environmental effects could induce dynamic perturbations of the
kind experienced by our unstable migrating BSs?}

One important point we would like to emphasise is that this
work is focused on the mergers of \emph{in-phase} BSs. However, as
first investigated in \cite{Palenzuela:2007dm}, it is well known
that \emph{off-phase} BSs exhibit significant deviation in the
merger dynamics from in-phase BS. Broadly speaking, BSs with different
phases exhibit \emph{repulsive} forces with the magnitude depending
on how off-phase they are \cite{Evstafyeva:2022bpr} -- this is due
to the fact that BSs are wave-like objects and can interfere either
constructively or destructively depending on their phases.  It is
unclear at the moment how these interference effects will affect
mergers of unstable or near-unstable BSs. As a guess to one of its possible effects in BSs, we note that
in off-phase mergers of real scalar Oscillatons
\cite{Widdicombe:2019woy}, it was shown that while off-phase
repulsion resists the formation of BHs from direct mergers, the
subsequent ``compression'' of \emph{individual} Oscillatons due to the repulsion
can cause sufficiently  large disruption such that it collapses into a BH on
its own. Thus, applying this logic to each BS, we anticipate that this might lead to a widening of the ``red dots
band'' say in Fig.~\ref{fig:appetizer}, although clearly there
is plenty of scope for further investigation in this direction.
Also, examining how waveforms change with the self-interaction parameter would provide a more complete perspective, and we leave this for future work.

Finally, any program to search for BSs in the observational data
stream requires the construction of waveform models or templates
akin to the \texttt{IMRPhenom}, \texttt{SEOBNR} or \texttt{NRSur}
approximants
for BH-binary mergers. Our results suggest that the GWs from the BS star merger
phase can exhibit significant differences in the GW amplitude as
compared to BH mergers; cf.~Fig.~\ref{fig:psi4_diagram}.
The GW amplitude, however, is essentially degenerate with the luminosity
distance and thus of little help for characterizing the nature of the 
source. This naturally raises the question {\it whether the GW signature of
the needle regime may exhibit more structural features in inspirals and/or
for other BS parameters like unequal mass ratios or non-zero spins.}
Depending on the longevity of BS perturbations, comprehensive binary
BS templates may need to account for these effects, perhaps using
phenomenological parameters to extend simple semi-analytic quadrupole
models (e.g.~Ref.~\cite{Toubiana:2020lzd}) with rigid mass distributions.
Given the complexity of the merger
physics, however, we suspect that surrogate models,
as for example constructed for Proca-star head-on mergers in  Ref~\cite{Luna:2024kof},
may provide the most efficient way forward.

\acknowledgments
This work has been supported by.
STFC Research Grant No.~ST/V005669/1,
NSF Grant Nos.~PHY-1626190 and PHY-2110594,
and
ExCALIBUR Hardware and Enabling Software
Grant Nos.~ST/X001393/1 and EP/Y028082/1.
We acknowledge support by the Cambridge Service for Data Driven Discovery
at the University of Cambridge and
Durham University through DiRAC Projects ACTP284 and ACTP and STFC capital Grants No.~ST/P002307/1 and
No.~ST/R002452/1, and STFC operations Grant No.~ST/R00689X/1.
We acknowledge support by the Texas Advanced Computing Center (TACC) and the San Diego Supercomputing Center (SDSC) through NSF
Grant No.~PHY-090003.
Computations were done on the CSD3 and Fawcett
(Cambridge), Cosma7 and 8 (Durham),
Stampede2 and Stampede3 (TACC), and Expanse (SDSC) clusters. EL and BG acknowledge computations resources provided by a DiRAC RAC15 grant ACTP316.
RC and TE have been supported by the Centre for Doctoral Training (CDT)
at the University of Cambridge funded through STFC. TE acknowledges Perimeter Institute for Theoretical Physics, supported by the Government of Canada through the Department of Innovation, Science and Economic Development and by the Province of Ontario through the Ministry of Colleges and Universities

\begin{figure}[t]
  \centering
  \includegraphics[width=0.85\linewidth]{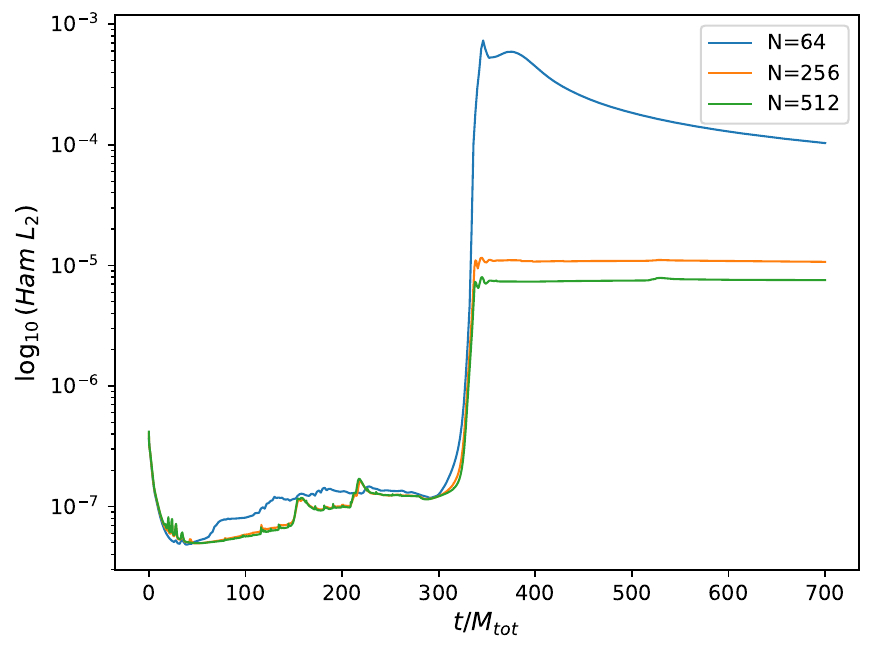}
  \caption{The $L_2$ norm of the Hamiltonian constraint as a
  function of time obtained for the
  BS binary configuration $\sigma_0=0.25$, $|\phi_{\rm c}|= 0.02$
  using different resolutions $N=64$, $N=256$ and $N=1024$.}
  \label{fig:ham}
\end{figure}
%
\appendix
\section{Constraint violations}
\label{app:ham}
The constraint violations observed in our simulations follow
the pattern exemplified for the Hamiltonian constraint $H$ of
the BS collision with parameters $\sigma = 0.25$, $|\phi_{\rm c}|=0.02$
in Fig.~\ref{fig:ham}. At early times, the constraint
damping of the CCZ4 formulation reduces the constraint violations
by an order of magnitude independent of numerical resolution.
During the remainder of the infall, up to about $t=300\,M$,
the constraint violations of the $N=256$ and $N=512$
simulations nearly overlap. We attribute this 
to the fact that the constraint violations have very low
magnitude and are initial-data dominated.
To test this hypothesis, we have
added a third simulation with lower resolution $N=64$. This indeed exhibits a
significant contribution to $H$ arising from the discretization error
with convergence of about second order.
During merger, the constraint violations increase substantially, as
expected for the formation of the extreme curvature region of a black hole
\footnote{Note that this effect is enhanced by the effectively
two-dimensional domain, which for purely geometric reasons attaches
relatively higher weight to grid points near the black hole as compared to a
three-dimensional grid.}. 
Post merger, we observe convergence between first and second order.

\section{Gravitational Wave Extraction}
\label{app:waveextraction}
Numerical relativity simulations in $3+1$ dimensions commonly use
the Newman-Penrose scalar $\Psi_4$ to extract gravitational
radiation. Here we follow the discussion of GW extraction in $d+1$
dimensions of Ref.~\cite{Cook:2016qnt} and apply it to our case of
$2+1$ evolutions ($d=2$). The outgoing gravitational radiation is
encoded in the the projections of the Weyl tensor $C_{A B C D}$
onto a null tetrad $\{k^A, l^A, m^{A}, \bar{m}^{A}\}$
\begin{equation} \label{eq:Psi4_GW}
  \Psi_4 = C_{ABCD} k^A \bar{m}^B k^{C} \bar{m}^D\,;
\end{equation}
here capital Latin indices run from $0$ to $d$.  In order to construct
the null tetrad, we first define an orthonormal basis $\{e_{(0)}^A,
e_{(1)}^A, e_{(2)}^A, e_{(3)}^A \}$, where $e_{(0)}^A$ is the unit
normal to the spatial hypersurfaces, $e_{(1)}^A$ is the unit normal
radial vector and $e_{(2)}^A, e_{(3)}^A$ are the angular vectors,
which are typically constructed via Gram-Schmidt orthonormalisation.
We then pick the tetrad consisting of ingoing and outgoing null
vectors, $k^{A}$ and $l^{A}$ respectively, and a complex null vector
$m^{A}$ and its conjugate $\bar{m}^{A}$,
\begin{align}
  l^{A} &= \frac{1}{\sqrt{2}}      
        \left(e_{(0)}^A + e_{(1)}^A \right), \\
        k^{A} &= \frac{1}{\sqrt{2}} \left(e_{(0)}^A - e_{(1)}^A \right), \\
  m^{A} &=\frac{1}{\sqrt{2}} \left(e_{(2)}^A     + ie_{(3)}^A \right), \\
        \bar{m}^{A} &=\frac{1}{\sqrt{2}} \left(e_{(2)}^A - ie_{(3)}^A \right).
\end{align}
Following the notation of Ref.~\cite{Cook:2016qnt}, we will denote
the off-domain components with a '$w$' index, so that in $2+1$ we have,
$e_{(3)}^{A} \rightarrow e_{(w)}^{A}$. It is then straightforward
to show that the outgoing gravitational radiation in Eq.~\eqref{eq:Psi4_GW}
can be re-expressed by the real quantity
\begin{align}
  \Omega_{(a)(b)} &= \frac{1}{4} (R_{0B0D}e_{(a)}^B
  e_{(b)}^D - R_{AB0D} e_{(1)}^A e_{(a)}^B e_{(b)}^D \nonumber \\
  &- R_{0BCD} e_{(a)}^B e_{(b)}^D e_{(1)}^C + R_{ABCD}
  e_{(1)}^A e_{(a)}^B e_{(b)}^D e_{(1)}^C),
\end{align}
where we have introduced early lower-case Latin indices $a, b, \ldots$
to span the angular directions (i.e.~2 and 3) and used the fact
that the GW signal is calculated in vacuum, where the Weyl and
Riemann tensors are the same. Therefore different $(a, b)$
indices correspond to different polarisation content. A direct
calculation results in the following relationship between
$\Omega_{(a) (b)}$ and $\Psi_4$ in $2+1$ dimensions,
\begin{align}
  \rm{Re}(\Psi_{4}) &= \Omega_{22} - \Omega_{ww} \\
  \rm{Im}(\Psi_{4}) &= - 2\Omega_{2w} = 0,
\end{align}
where $\Omega_{22}$ and $\Omega_{ww}$ can be obtained directly from
Eqs.(4.22)-(4.35) of Ref.~\cite{Cook:2016qnt}.

\bibliography{newuli2}

\end{document}